# Recent progress and challenges in magnetic tunnel junctions with 2D materials for spintronic applications

Lishu zhang[ab], Jun Zhou[b], Hui Li*[a], Lei Shen*[c], Yuan Ping Feng*[bd]

[a] *Key Laboratory for Liquid-Solid Structural Evolution and Processing of Materials, Ministry of Education, Shandong University, Jinan 250061, China*
[b] *Department of Physics, National University of Singapore, 2 Science Drive 3, Singapore 117542, Singapore*
[c] *Department of Mechanical Engineering, National University of Singapore, 9 Engineering Drive 1, Singapore 117542, Singapore*
[d] *Center for Advanced 2D Materials, National University of Singapore, 6 Science Drive 2, Singapore 117546, Singapore*

* Corresponding author: lihuilmy@hotmail.com (H.L.); shenlei@nus.edu.sg (L.S.); phyfyp@nus.edu.sg (Y.F.);

## Abstract

As Moore's law is gradually losing its effectiveness, developing alternative high-speed and low-energy-consuming information technology with post-silicon advanced materials is urgently needed. The successful application of tunneling magnetoresistance (TMR) in magnetic tunnel junctions (MTJs) has given rise to a tremendous economic impact on magnetic informatics, including MRAM, radio-frequency sensors, microwave generators and neuromorphic computing networks. The emergence of two-dimensional (2D) materials brings opportunities for MTJs based on 2D materials which have many attractive characters and advantages. Especially, the recently discovered intrinsic 2D ferromagnetic materials with high spin-polarization hold the promise for next-generation nanoscale MTJs. With the development of advanced 2D materials, many efforts on MTJs with 2D materials have been made both theoretically and experimentally. Various 2D materials, such as semi-metallic graphene, insulating h-BN, semiconducting $MoS_2$, magnetic semiconducting $CrI_3$, magnetic metallic $Fe_3GeTe_2$ and some other recently emerged 2D materials are discussed as the electrodes and/or central scattering materials of MTJs in this review. We discuss the fundamental and main issues facing MTJs, and review the current progress made with 2D MTJs, briefly comment on work with some specific 2D materials, and highlight how they address the current challenges in MTJs, and finally offer an outlook and perspective of 2D MTJs.

**Keywords**

Magnetic tunnel junctions, spintronics, 2D materials, spin-orbit torque, magnetoresistance

## Contents

## 1. Introduction

The conventional silicon-based metal-oxide-semiconductor devices which work on the manipulation of charges (one degree of freedom of electrons) will come to an end in the near future due to more fundamental issues. Exploring advanced information technology with high-speed operation and low-energy consumption to replace existing silicon-based technologies is urgently demanded. So far, many strategies have been proposed, like nanoelectronics,[1-4] molecular electronics,[5] spintronics,[6-9] and quantum information technologies[10-12]. Among them, spintronics has exhibited tremendous potential and thus attracted lots of attentions. Spintronics which is based on manipulation of spins (another degree of freedom of electrons) have promise to integrate memory technology at the heart of information processing units such as classical and neuromorphic, which would be a big change in how architectures are designed towards in-memory computing (currently memory and logic are separated layers that need to communicate). In addition, spintronics is more compatible with conventional electronics, compared to other strategies, so that many techniques applied in traditional electronics can be extended to spintronics.

Even though information is processed using spin, it is desirable to manipulate spin or switch magnetization using electrical means with magnetoresistive materials. Magnetoresistive devices are constructed based on the magnetoresistive materials which in general exhibit a change in resistance with the application of a magnetic field. Early in its development, magnetoresistance devices made use of the anisotropic magnetoresistance (AMR) effect. Recently, such devices are mainly based on giant magnetoresistance (GMR) effect, or tunneling magnetoresistance (TMR). The discovery of magnetic tunnel junctions (MTJs) at room temperature dramatically increased the storage density.[13, 14] If a MTJ has TMR over 100%, it can be used to make not only magnetic field sensors[15] and reading heads of hard drives,[16] but also magnetoresistive random access memories (MRAM)[17]. The more recent demonstration of spin transfer torque (STT) [18] and spin-orbit torque (SOT) effect [19, 20] makes MTJs more valuable for manufacturing multitudinous spintronics, including MRAM,[21-23] radio-frequency sensors,[24, 25] microwave generators[26] and even artificial neuromorphic networks[27]. STT allows change of the magnetization direction of a material by a spin-polarized current. By passing a current through a thick magnetic layer (the “fixed layer”), one can produce a spin-polarized current. If this spin-polarized current is directed into a second, thinner magnetic layer (the “free layer”), the angular momentum of charge carriers (such as electrons) can be transferred to this layer, changing its magnetization orientation. This can be used to excite oscillations or even flip the orientation of the magnet. Thus, the different resistive states, a low-resistance parallel (P) magnetic configuration and high-resistance state, anti-parallel (AP) magnetic configuration), can be realized to represent the ‘0’ and ‘1’ state in the STT memory respectively.[28-30] Thanks to voltage-depended switching ability for STT principle, the different logic gates can be reconstructed by the same MTJ structures with only one single-cycle operation. This promising computing hardware application proves MTJs have great potential to be applied in more aspects in the future. SOT is another approach of magnetization switching, an interconversion of charge and spin current, which is a promising phenomenon that can be used to improve the performance of MRAM devices[21-23]. The SOT effect has been achieved in heavy metal [31] and 2D topological insulator (TI) systems[32]. It essentially requires two functional layers, namely, one ferromagnetic (FM) and one nonmagnetic layer with large spin-orbit coupling (SOC). The latter is to accumulate spin charges and inject it into the adjacent FM layer. The spin current then exerts a torque on the magnetic moment of the FM layer and revert it with an angle. The switching efficiency strongly depends on the strength of SOT. Furthermore, a large spin polarization in the nonmagnetic layer is necessary for efficient spin injection. Thus, heavy metals and 2D TIs with large spin Hall angles, such as Bi [33] and 2D-TI α-Sn,[34] are used for SOT switching. Much progress has been made in SOT magnetic switching with heavy metals [35, 36]. Recently, TIs, such as $Bi_2Se_3$, have attracted attention due to their spin-momentum locking property which in principle is able to achieve an efficient SOT switching, even though the switching process needs to be further improved in terms of the switching hysteresis and its completeness.[37]

Advantages MTJs devices include low-power consumption with high processing speed, non-volatility, metal–oxide–semiconductor technology compatibility and high integration density.[38,39] The most common materials used to fabricate MTJs are ferromagnetic metals and alloys such as Fe and CoFeB, Heusler alloys and dielectrics like MgO and $AlO_x$.[40-43] What cannot be ignored is that remarkable breakthrough has also been achieved in 2D materials synthesis and MTJs begin to be created based on them.[44] Due to its low dimensionality and quantum nature, the use of 2D materials adds many unique features into MTJs such as flexibility, and extremely high scaling.[45, 46]

First-principles calculations based on density functional theory (DFT) have been the most widely used method in theoretical studies of 2D materials and their applications in devices such as MTJs. It can be said that the rational design of high-performing MTJs would always face challenges in experiments without proper theoretical guidance, such as in the development of (Co)Fe/MgO/(Co)Fe MTJs. Compared to other theoretical and computational methods, the first-principles method does not require empirical parameters and experimental inputs, which makes it an ideal method for studying new materials and their heterostructures. DFT calculations are also valuable in predicting materials and device behaviors under extreme conditions that are difficult to achieve experimentally.

DFT was proposed by Hohenberg and Kohn in 1964.[47] And at the very next year Kohn and Sham launched its primary fulfillment.[48] DFT has become a convincing quantum simulation method in exploring the electronic structure of many systems through the use of the electron density as the fundamental variable instead of the electron wave function. The first MTJ based on Fe/MgO/Fe was proposed by DFT calculations,[49] and was subsequently demonstrated experimentally[40, 50]. Now, the Fe/MgO/Fe-based MTJs are the main components in the reading head of the hard disk in personal computers. Besides, in searching for the magnetism of 2D materials and predicting new structures, first-principles calculations has also been used as a powerful tool for providing theoretical guides for experimental exploration.

Even though DFT has been very successful in studying and predicting new materials, it is still a computationally expensive method. Despite of significant improvements in recent decades, it is still difficult to incorporate all of the experimental "real world" subtlety. Concerning calculation of transport properties, many systems beyond MgO have been initially predicted by DFT to lead to high spin polarizations (with similar symmetry arguments), but experimentally their performance so far failed to match that of MgO. As a result, direct comparison between computational prediction and experimental measurement for quantities such as MR ratio is non-trivial. Considering also the fact that there have been many computational studies on MTJs but a limited experimental realization of the predicted structures, here we mainly focus on computational works in this review, but also discuss available related experimental works. We begin with a brief introduction to spintronics and conventional MTJs. And then we discuss the current status on MTJs, highlighting the problems encountered and challenges. This is followed by the advantages of 2D materials in solving those MTJs challenges. The rest of this review is organized as follows. In **Sec. 4**, we review the recent progress of 2D-materials-based MTJs. This is organized into four subsections based on the key issues in 2D MTJs, i.e., targeting at spin polarization (**Sec. 4.1**), spin injection (**Sec. 4.2**), spin manipulation (**Sec. 4.3**) and stability (**Sec. 4.4**), respectively. We finally conclude and offer an outlook and perspectives in **Sec. 5**.

### 1.1 Spintronics

Conventional electronic devices have one thing in common, that is, they rely on the electronic transport in semiconductor materials such as silicon. With the size and function of silicon-based electronic devices reaching the limit, further downscaling of silicon based electronic devices becomes impossible. New concepts are required for future electronic devices which should also meet certain requirements such as low power operation. To this respect, it is noted that the energy scale of spin dynamics is typically many orders of magnitude smaller than that of charge dynamics, and low power electronics operation can thus be achieved in spintronic devices. Spintronics has become as a rapidly developing field under this background. Spintronics is based on the manipulation of spin of electrons to store, encode and transmit data.

In spintronics, information is first marked as up spin or down spin; and the spin-carrying electrons are transported along a path; and finally at a final point, the spin information is read. The conduction electrons' spin orientation needs to sustain for several nanoseconds, in order for them to be used in electrical circuit and chip. Transporting current through a ferromagnetic material and transmitting the spin-polarized electrons to the receiver is a common method to generate spin-polarized current. The successful implementation of spintronic devices and circuits (**Fig. 1**) relies on the realization of six elementary functionalities: spin–orbital torque, spin detection, spin transport, spin manipulation, spin-optical interaction and single spin device. Spin–orbital torque is induced through the spin–orbital interaction in FM/heavy metal bilayers by flowing an in-plane electrical current.[51] The spin detection includes detection of circularly polarized light,[52] transient Kerr/Faraday linearly polarized light rotation,[53] spin Hall voltage,[54, 55] electric resistance change,[56] and tunneling-induced luminescence microscopy,[57] which has been the most convenient method from device perspective and applications till date. Spin transport is expected to offer low-loss spin channels, which can provide long-distance propagation of spin signals and enable more operations of spin signals.[58, 59] Spin manipulation is required to achieve more functionality of spintronic devices like that in electronic devices.[60-62] Spin-optical interactions include spin-Hall effects in inhomogeneous media and at optical interfaces, spin-dependent effects in nonparaxial (focused or scattered) fields, spin-controlled shaping of light using anisotropic structured interfaces (metasurfaces), and robust spin-directional coupling via evanescent near fields.[63] The single-spin, singlet, and polarized phases of a quantum dot allow different currents to flow through the dot. The spin state of the dot is controlled either by adding electrons or by tuning the magnetic field, and thus a prototype single-spin transistor is produced. [64, 65]

At present, a variety of spin electronic devices based on different mechanisms of spintronics have been studied and designed. Several applications are also highlighted in **Figure 1**. In a SOT-driven device, the heavy metal induces strong SOC and thus generate a SOT-driven switching. In a spin logic device, by defining bistable magnetizations of electrodes along the easy axis as the *input* logic ('1' and '0') and the as-detected current as the logic *output*, Boolean operations can be achieved. The spin electrons transport from one FM layer to another FM layer, by passing through a barrier, high and low spin current can be achieved by the magnetic alignment of the two electrodes. As such, this kind of devices, called MTJs, can be used as a memory device to store information even under the power-off state (nonvolatile). Spin field effect transistors (FET) was proposed by Datta and Das first.[66] It is based on manipulation of electron spin during transport driven by an electric field in semiconductors. This device works similarly to a charge-based transistor. A spin current is injected into the channel material from a FM electrode (source), in which spin polarization is electrically manipulated by a gate voltage (or other means), and finally spin polarization is detected at the drain. In a spin light emitting diode (LED), when the spin-polarized electron is injected, it recombines with a hole and emits a circularly polarized photon which is used to assess the polarization of the injected spin. When a spin polarized electron is injected into a quantum dot, the spin state of the quantum dot can be changed and can be controlled by a gate voltage. All these spintronic devices have lower power consumption, lower cost, more stable and excellent performance in high-capacity storage than traditional electronic devices whose operation principle is only based on charge. Therefore, spintronic devices will play a great role in the next generation of electronic information science and technology.

In the past several years, three important focus areas of spintronics research have been explored by scientists: 1) fabricating nanoscale structures including new magnetic materials, hybrid heterostructures, and functional materials; 2) studying the spin effect including spin injection, transport and detection; and 3) improving the performance of MTJ-based devices.

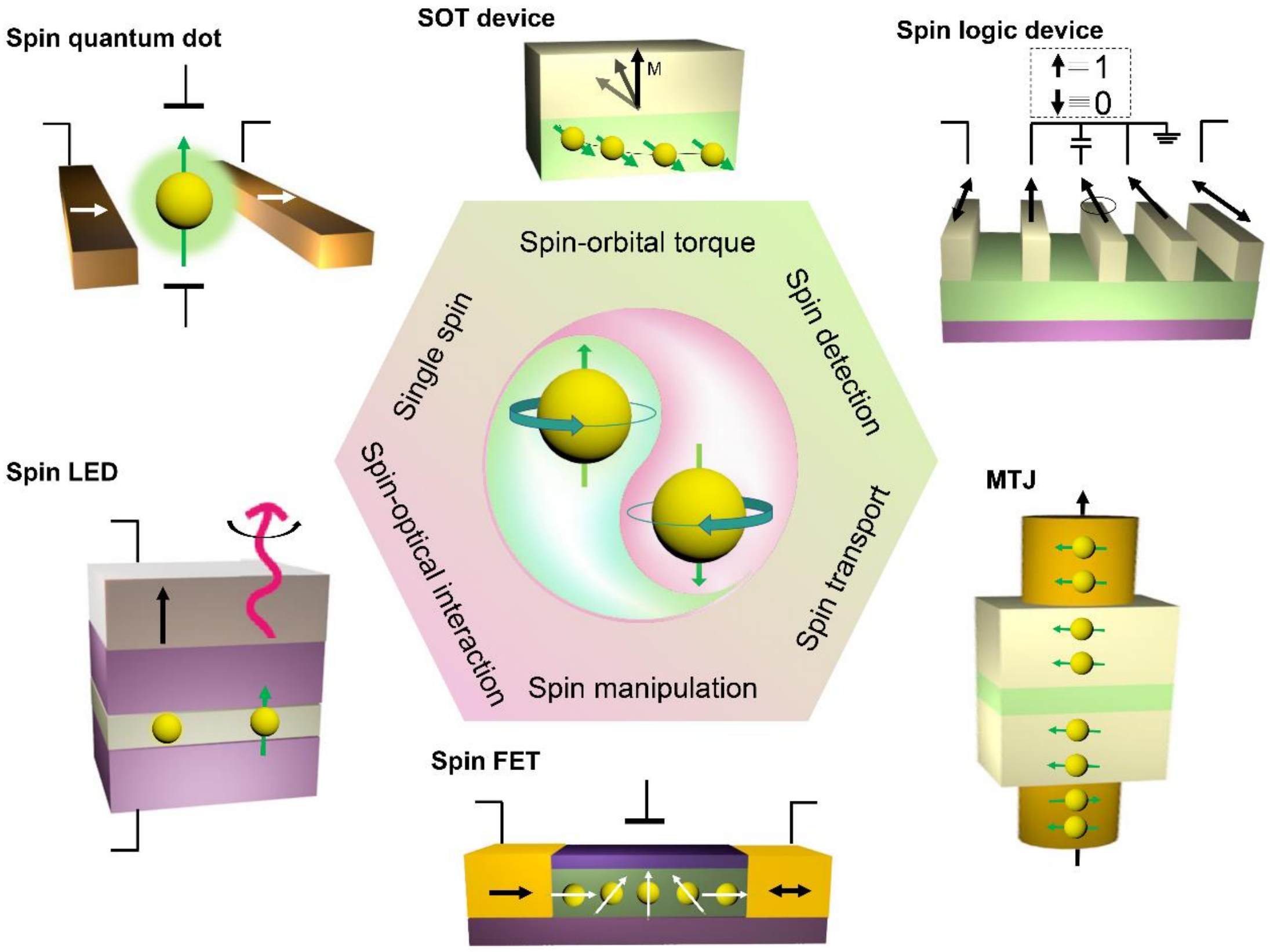

**Fig. 1** Overview of some spintronics devices. Key aspects, including physical effects, elementary functionality and applications, are schematically illustrated. The elementary functionalities include spin–orbital torque, spin detection, spin transport, spin manipulation, spin-optical interaction and single spin. The corresponding applications of these functionalities include the SOT device, spin logic device, MTJ, spin FET, spin LED, and spin quantum dot device. In this review, we focus on the magnetic tunnel junction.

## 1.2 MTJs

Under the well-developed knowledge on how to manipulate spins,[67-74] one can generate state-of-the-art spintronics devices with desired properties. Thus, it is vital to explore the application possibilities of spintronic effects in order to achieve more promising spintronics devices. Such electronic devices have made a big impact on computer technology through achieving higher and higher information storage in hard desk drives as well as faster and faster reading speed of data in RAMs. MTJ is one of the most important forms in spintronics applications as mentioned in **Fig.1**. In this section, we will introduce MTJs.

A basic MTJ consists of two ferromagnetic layers separated by a thin insulating layer, as schematically shown in **Figs. 2 (a)** and **(c)**. The tunneling conductance or resistance of such a device depends on whether the magnetizations of the two electrodes are parallel or antiparallel. If $R_P$ and $R_{AP}$ are the resistance in the parallel and antiparallel state, the tunneling magnetoresistance (TMR) ratio is given by[75]

$$\mathrm{TMR} = \frac{R_{AP} - R_P}{R_P} = \frac{2P_1P_2}{1 - P_1P_2}$$

where $P_1$ and $P_2$ are the spin polarization of the two electrodes. The origin of TMR arises from different density of states (DOS) for spin up and spin down electrons as shown in **Figs. 2 (c)** and **(d)**. Because electron spins are preserved during the transport, each type of spin can only tunnel into the subband of the same spin. Therefore, the tunnel current is high (or resistance is low) when the magnetizations of the two electrodes are parallel due to the matching DOS on both sides [**Fig.2(b)**], and that in the antiparallel state is low (or resistance is high) [**Fig.2(d)**], even though this may change depending on spin selection at the interface. It is worth noting that although it is insignificant for small TMR, for large negative TMR, the resistance variation is sometimes normalized preferentially by $R_{\mathrm{AP}}$, so as to obtain the same value as positive TMR, and thus being comparable in absolute value.[76]

TMR at room temperature was first demonstrated by Miyazaki [77] and Moodera [78]. Immediately after that the TMR ratio was risen rapidly to 81% in a $Co_{0.4}Fe_{0.4}B_{0.2}$ (3)/Al (0.6)-$O_x$/$Co_{0.4}Fe_{0.4}B_{0.2}$ (2.5) (thickness in nm) MTJ at room temperature.[79] Subsequently, TMR ratio as large as 604% was achieved in MgO based MTJ, $Co_{0.2}Fe_{0.6}B_{0.2}$ (6)/MgO (2.1)/$Co_{0.2}Fe_{0.6}B_{0.2}$ (4) (thickness in nm) at room temperature.[80] The dramatic increase in MR ratio, compared to that of its predecessor, GMR devices, led to the domination of MTJs in magnetic data storage industry.

The first successful application of MTJ was demonstrated in computer read head technology with $Al_2O_3$ barrier and MgO barrier MTJs. The magnetic recording density in the hard disk drive increased considerably compared to traditional devices.[81-85]Another MTJ application is to develop the MRAM which exceeds the density of Dynamic RAM (DRAM), speed of static RAM (SRAM) and non-volatility of flash memory. Moreover, these nanoelectronics generate less heat and operate at lower power consumption.

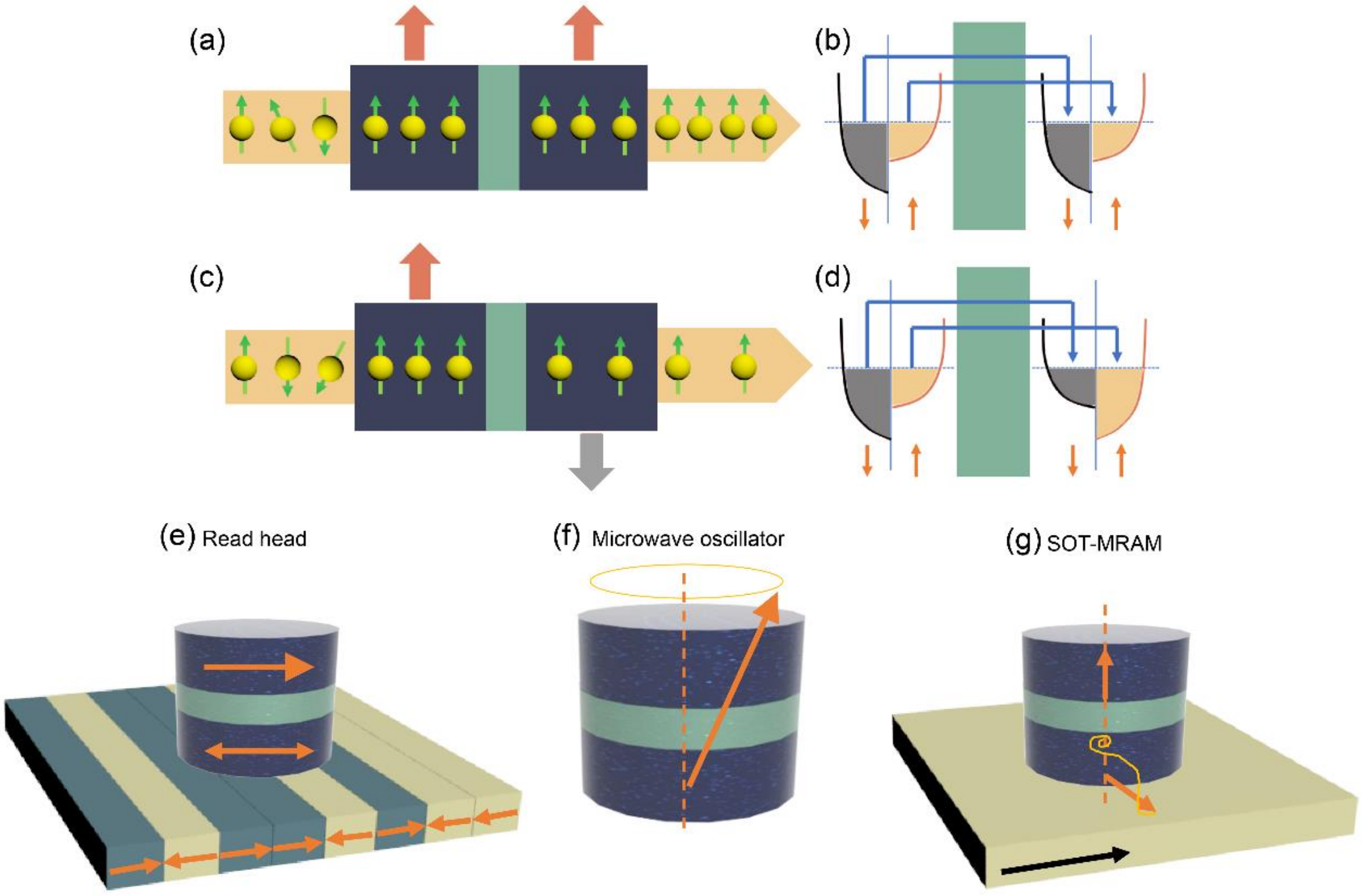

**Fig. 2** (a-b) Schematic diagram of MTJs in P configuration and corresponding band diagram. (c-d) Schematic diagram of MTJs in AP configuration and corresponding band diagram. (e-g) The exampled applications for MTJs, i.e., the read head, microwave oscillator and MRAM.

Because there is no or almost no interlayer coupling between the two ferromagnetic layers in MTJs, only a small external magnetic field is needed to reverse the magnetization direction of one ferromagnetic layer, thus realizing a huge change in tunneling resistance. Therefore, MTJs have much higher magnetic field sensitivity than metal multilayer films. At the same time, MTJs have high resistivity, low energy consumption and stable performance. All in all, MTJs act as one of the most important spintronics applications, which can use as a key component in many spintronics devices, such as, the read head of hard disk drives, microwave oscillator and MRAM, as shown in **Figs. 2 (e-g)**. Recent studies have demonstrated that magnetization in the free layer of an MTJ can be switched by STT or SOC, even though in practice it is not so easy due to highly spin polarized current density required.

To fabricate an MTJ with a giant TMR is crucial for practical applications. With the development of nanotechnology, there are more and more ways to construct junction structures. For the preparation methods in the laboratory, methods such as molecular beam epitaxy (MBE),[86] magnetron sputtering,[87] electron beam evaporation[88] and chemical vapour deposition (CVD),[89] are often used. In industry, methods used to prepare micron, sub-micron and nano magnetic tunnel junction, magnetic tunnel junction array, TMR magnetic read-out head and MRAM include lithography, electron beam exposure, ion beam etching, chemical reaction etching, focused ion beam etching, etc. Among them, lithography combined with ion beam etching is the preferred process with low cost and mass production in micromachining process. Generally, all MTJs consisted of FM layers and an insulator layer. The most common ways to fabricate the FM layer in the past years is sputter deposition (magnetron sputtering and ion beam deposition).[90-92] The magnetic alignment and thickness are the key parts of MTJ fabrication in the experiment. A better method to fabricate insulating layer always keeps forging ahead. For example, ion beam oxidation,[93] glow discharge,[94, 95] plasma,[96] atomic-oxygen exposure[97] and ultraviolet-stimulated oxygen exposure[98] have been used as alternate ways for the insulator-layer deposition. In terms of preparation and processing, the issues about the control of the oxide barrier and the interfaces, the shielding tolerance, the thermal stability, and the robustness of the lifetime of the device need to be solved urgently.

## 2. Current status and issues of MTJs with 3D materials

With the rapid development of MTJs, the TMR value of MTJs increases rapidly in the last few years, and quickly approaches the theoretical value. However, at present, despite of extensive studies and much progress has been made, there are still many problems and challenges that need to be understood and addressed to improve the efficiency, performance and stability of MTJs. For example, one of the important issues is to control the quality of the interface between ferromagnetic layer and barrier layer.[76] The effect of the interface bonding on magnetoresistive properties must be considered,[99, 100] because it determines the effectiveness of transmission of electrons with different orbital properties (and/or symmetry) through the interface, and electrons with different orbital properties carry unequal spin polarization.[101] Therefore, the interface bonding has profound effects on conductance.[102] For this aspect, layered 2D materials which is connected by van der Waals (vdW) force without chemical bonding can avoid these problems. And also, it is known that a long spin lifetime in the nonmagnetic (NM) materials and an efficient polarization of the injected spin are required. Fert *et al.*[100] show that introducing a spin dependent interface resistance at the FM/NM interfaces can solve the problem of the conductivity mismatch between FM and NM materials. They find a significant magnetoresistance can be obtained if the junction resistance at the FM/NM and NM/FM interfaces is chosen in a relatively narrow range depending on the resistivity, spin diffusion length and thickness of NM. However, introducing 3D tunneling barrier such as $Al_2O_3$ indeed has effect on improve spin injection efficiency[103, 104], but new issues such as pinholes and clusters emerge[105]. And using 2D materials like h-BN which owns well defined interface contact and less defects to act as tunneling barrier can avoid these new issues perfectly[106]. In addition, the spin-dependent electronic structure of electrodes,[107] the symmetry selection rules that are known to control TMR in MTJs with electrodes and crystalline tunnel barriers,[50] the role of the tunnel barrier layer and its electronic structure[49, 108] are all important issues and deserve serious attention. Moreover, the diffusion and oxidation process of elements in barrier layer is another puzzling issue in the growth process. Taking CoFeB/MgO/ CoFeB MTJs as an example, Burton *et al.* found that B atoms in the crystalline CoFeB electrodes tend to migrate to the interface which leads to a decrease in the TMR ratio due to a significant suppression of the majority-channel conductance through states of $\Delta_1$ symmetry.[109] Similarly, the diffusion and oxidation process also have an influence on tunneling process.[110] Another thing is the tunneling mechanism of barrier layer.[111] MgO systems show much larger TMR than that of traditional $AlO_x$ systems.[112-114] However, the tunneling mechanism in single-crystalline or textured MgO barriers is quite different from traditional $AlO_x$ amorphous barrier materials.[111] In addition, perpendicular magnetic anisotropy (PMA) of out-of-plane magnetized MTJs aroused a lot of attention, because in-plane anisotropy only yields a typical anisotropy field in the 100–200 Oe due to shape anisotropy, while PMA can yield an effective anisotropy field of several kOe.[115] In order to realize the miniaturization of devices, it is urgent to develop processing technology on atomic scale. However, controlling thickness of metal oxide tunnel barriers is hard and challenging.[116, 117] 2D materials have the natural advantage of ultra-thin to atomic scale, which can avoid some troubles in processing. And it also needs to develop devices working at room-temperature. But most of the existing FM materials own too low Curie temperature ($T_C$) caused the need to search for room temperature FM. Last but not least, the origin and the influence of different layer thicknesses on the transport properties are also important. Understanding and resolve these issues will greatly promote the progress of MTJs in theoretical and practical applications. The discovery of 2D materials and their heterostructures provides a new playground for MTJs. Some of the 2D materials may offer promising routes to resolve some of these issues with their unique properties like sharp interfaces, natural and tunable van de Waals insulating gap, layer-by-layer control of the thickness, high PMA, the potential for a diffusion barrier (thermal stability), and even provides the possibility of new functionalities such as spin filtering. Under this background, this review discusses and summarizes around the following four main problems, as shown in **Fig. 3**: (1) spin polarization, (2) spin injection, (3) spin manipulation, and (4) spin stabilities.

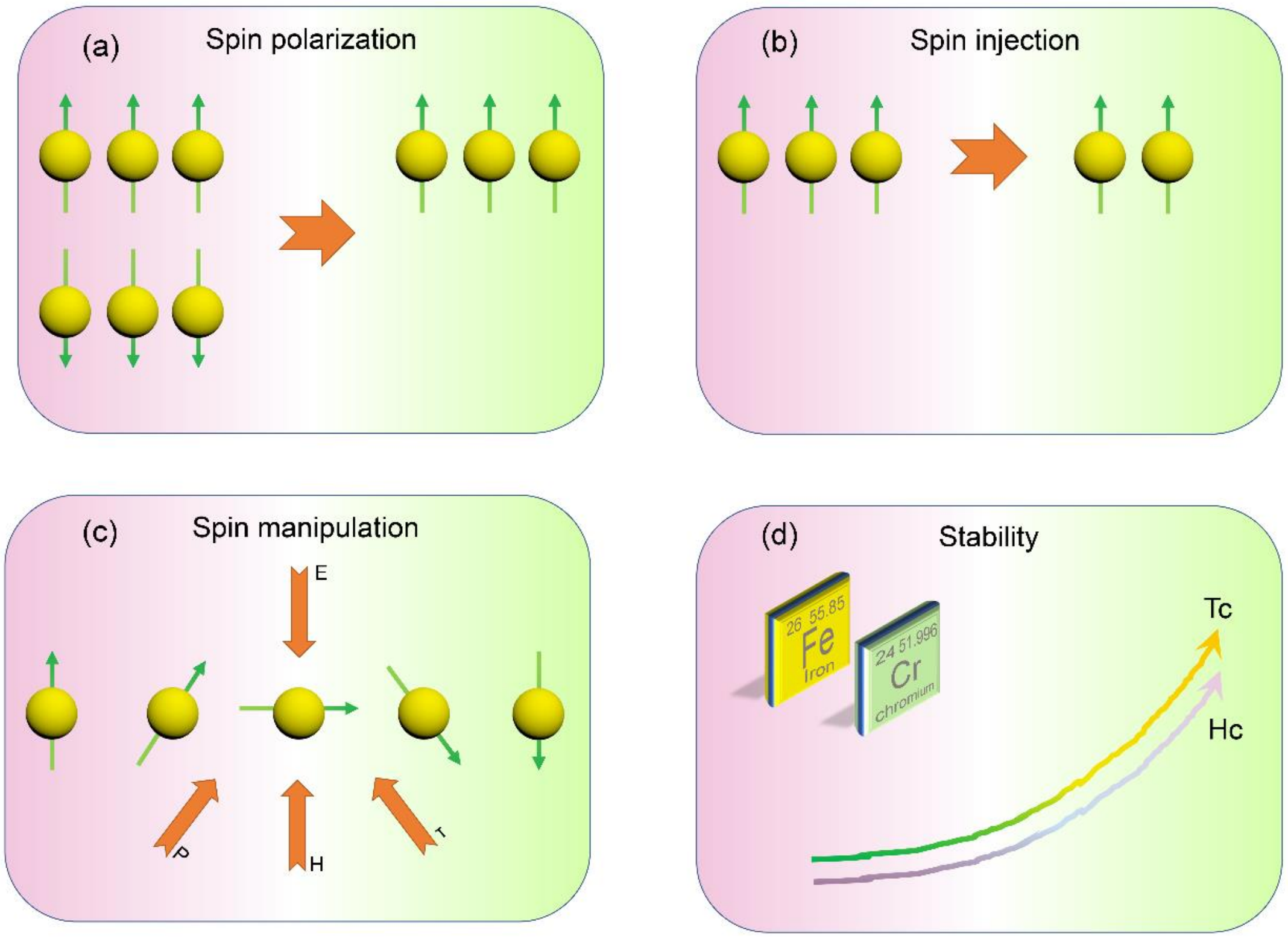

**Fig.3** Typical problem-oriented designs via MTJs.

## 3. 2D materials

When thin down the layered materials to their physical limits, they exhibit novel properties which is different from their bulk counterpart. Thus, these materials are specifically referred to as "2D materials". In other words, 2D materials refer to materials in which electrons can only move freely in two dimensions (in a plane) on the non-nanometer scale (1-100 nm). 2D materials emerged with the successful separation of graphene, a single atomic sheet of carbon atoms with a bonding length of 1.42 Å, by Geim's team in Manchester University in 2004.[118] Graphene is also regarded as the most widely studied 2D material. It has been well-known that the pristine graphene is a unique 2D hexagonal structure with zero-bandgap and semi-metallic property, which is an important allotrope of carbon. Due to 2D materials' various crystal structures and physical properties, many other 2D materials beyond graphene are also undergoing a lot of research work, including semiconductors (e.g., transition metal dichalcogenides such as $MoS_2$), insulators (e.g., h-BN), superconductors (e.g., $NbSe_2$), and magnets (e.g., $Fe_3GeTe_2$). These great advances have expanded 2D nanodevices.

The research on MTJs has been committed to continuously improve the high TMR ratio all the time. However, in the process of further improving the performance and reducing the size, more and more challenges emerge. Some of the 2D materials may offer promising routes to resolve some of these issues with their unique properties, and even provides the possibility of new functionalities such as spin filtering.

### 3.1 Theoretical prediction and experimental synthesis

Recently, more and more 2D materials were predicted theoretically and some of them have been synthesized experimentally. Take *2Dmatpedia* database[119] as an example, this database contains 6,351 2D materials at present, of which 1500 materials are magnetic based on the spin-polarized DFT calculations (the total magnetic moment of the material is large than 0.5 μB). The known 2D magnetic materials are summarized in **Fig. 4** [120] and they are divided into different categories according to whether they are synthesized experimentally or predicted theoretically and their properties. By combining 2D materials with different properties from this table as heterojunctions, rich 2D MTJs can be constructed.

In order to make 2D magnetic devices, the long-range magnetic order in 2D magnetic materials is necessary. It is known that adjacent atomic moments or spins are coupled through an exchange interaction in a lattice, leading the magnetic order of materials.[121] The magnetism depends on the lattice dimensionality or crystal structure, as well as the spin dimensionality of the system. The *uniaxial anisotropy* is able to sustain long-range magnetic ordering, which has been experimentally observed in several magnetic 2D materials recently. It is worth noting that 2D Ising kind of behavior has been reported in neutron scattering experiments in layered materials much earlier.[122] The existence of magnetism down to monolayers in several magnetic 2D materials has been established very recently. Intrinsic magnetic order in 2D layered $Cr_2Ge_2Te_6$ was found at low temperatures in 2017.[123, 124] Almost the same time, the ferromagnetic order was also found in monolayer $CrI_3$ up to 45 K. Subsequently, several magnetic materials including $VSe_2$ and $Fe_3GeTe_2$ have been found at room temperature. These findings offer a new opportunity to manipulate the spin-based devices efficiently in the future.[125, 126]

| | | | |
|---|---|---|---|
| Chalcogenides | $Cr_2Ge_2Te_6$, $Cr_2Si_2Te_6$, $Fe_3GeTe_2$, $VSe_2^*$, $MnSe_x^*$ | $Fe_2P_2S_6$, $Fe_2P_2Se_6$, $Mn_2P_2S_6$, $Mn_2P_2Se_6$, $Ni_2P_2S_6$, $Ni_2P_2Se_6$, $CuCrP_2Se_6^*$, $AgVP_2S_6$, $AgCrP_2S_6$, $CrSe_2$, $CrTe_3$, $Ni_3Cr_2P_2S_9$, $MnBi_2Te_4^*$, $MnBi_2Se_4^*$ | $CuCrP_2S_6$ |
| Halides | $CrI_3^*$, $CrBr_3$, $GdI_2$ | $CrCl_3$, $FeCl_2$, $FeBr_2$, $FeI_2$, $MnBr_2$, $CoCl_2$, $CoBr_2$, $NiCl_2$, $VCl_2$, $VBr_2$, $VI_2$, $FeCl_3$, $FeBr_3$, CrOCl, CrOBr, CrSBr, $MnCl_2^*$, $VCl_3^*$, $VBr_3^*$ | $CuCl_2$, $CuBr_2$, $NiBr_2$, $NiI_2$, $CoI_2$, $MnI_2$<br>α-$RuCl_3$ |
| Others | $VS_2$, $InP_3$, GaSe, GaS | $MnX_3$ (X = F, Cl, Br, I), $FeX_2$ (X = Cl, Br, I), MnSSe, $TiCl_3$, $VCl_3$ | SnO, GeS, GeSe, SnS, SnSe, GaTeCl, CrN, $CrB_2$ |

**Fig. 4** 2D magnetic materials library. In this diagram, the gray line below lists theoretically predicted vdW ferromagnets (left), half metals (center), and multiferroics (right), respectively. And others above are 2D magnetic materials have been experimentally confirmed. Among them are bulk ferromagnetic vdW crystals (green-colored), bulk antiferromagnets (orange-colored), bulk multiferroics (yellow-colored) and a-$RuCl_3$ (a proximate Kitaev quantum spin liquid)[127] is colored by purple. *Reproduced with permission from Gong et al., Science **363**, 6428 (2019). Copyright 2019 The American Association for the Advancement of Science.*

### 3.2 Advantages of 2D materials in solving current 3D MTJs problems

In **Sec. 2**, the potential problems of 3D MTJs when they are scaling down to the nano level are introduced. Various 2D materials with natural monolayers are now available through the large scaled CVD growth. It is thus naturally to seek high-performing, flexible and stable MTJs tunnel barriers based on 2D materials and their heterojunctions. 2D materials provide a reliable solution to the problems in the manufacturing of high-performance MTJs through the layer-by-layer control of the thickness, sharp interfaces, and high PMA. In this direction, low resistance area products, strong exchange couplings across the interface, and high TMR in MTJs were predicted and synthesized.

## 4. 2D-materials-based MTJs

As discussed above, 2D materials are expected to offer solutions to some of the challenges when further improving the performance and reducing the size of MTJs. In this section, the current development status of 2D MTJs around the current problems in the development of conventional MTJs will be reviewed. Following each problem and challenge, possible solutions and future development directions are discussed.

### 4.1 Targeting high spin-polarization

The generation, transport and detection of spin current in MTJs are three key parts to integrate spin into existing electronics successfully. In this section, we focus on the current works which target at improving spin polarization in MTJs with 2D materials. The spin polarization is the most important factor for governing TMR performance as a high spin-polarized current is essential for high magnetoresistance. It is known that the subtle offset between two spin channels causes the net spin polarization, which is greatly affected by atomic, electronic and magnetic structures of the system. A straightforward way to have high spin polarization is to use *half-metallic* magnetic materials, in which the Fermi level only crosses one spin channel, resulting in 100% spin polarization. Thus, half-metallic materials become good candidates for MTJ devices. Using them as electrodes, 100% spin-polarized currents under a bias voltage may be generated inMTJs with high TMR. However, the MTJs with half-metallic-materials do not show very high TMR as expected from the materials point of view [cite reference here]. It is because of the complicated geometry structure in MTJ devices made of several different materials and nonequilibrium electronic structure under bias.

From the experimental aspect, with the fast development of CVD technology recently, MTJs using 2D materials such as $MoS_2$[128], graphene[129, 130] and boron nitride (BN)[131] as nonmagnetic spacer have been fabricated successfully. Their van de Waals interface is expected to overcome the disordered interface between two 3D bulk materials. However, their reported magnetoresistance is quite low, which are undesirable. This was attributed to the use of permalloy electrodes, such as Fe, Ni and Co, injecting current with a relatively low spin polarization. In addition, the inherent properties of these materials also hinder the performance of MTJs. Taking the Co-Fe system as an example, current-induced switching in FeCoB/MgO requires intense current densities to overcome the large Fe Gilbert damping.[132] Thus, it is important to search different 2D materials stacks with other electrodes, which can offer better opportunities to implement such new technologies. High spin-polarized Heusler alloys, a large family of ternary compounds,[133] appear as promising candidates of electrodes. For example, most early researches on $MoS_2$-MTJs with permalloy electrodes show relatively low TMR. Adopting the electrodes of $Fe_3Si$, a Heusler alloy with a lower Gilbert damping parameter and a higher saturation magnetization, Rotjanapittayakul *et al.*[134] reported a large TMR in $Fe_3Si$-$MoS_2$ MJTs. It is because of the similar lattice to $MoS_2$, small Gilbert damping and high Curie temperature of $Fe_3Si$. A small Gilbert damping parameter leads to a potentially low critical current density for STT switching. The $T_c$ of $Fe_3Si$ is large, above 800 K, and the spin-polarization at low temperature (~45%)[135] is larger than that of Co (~34%) and Ni (~11%)[136]. In addition, Wu *et al.*[137] preform a research on the ferromagnetic $Fe_3O_4$ electrodes in $Fe_3O_4/MoS_2/Fe_3O_4$ MTJs. A clear large TMR phenomenon appears below 200 K temperature. They also performed first-principles calculations and found that $Fe_3O_4$ keeps high spin-polarized electron band at the interface of the $MoS_2$ and $Fe_3O_4$. This calculation provides a clear and deep physical explanation on how the TMR phenomenon appears in their experiment.

By using the DFT combining with non-equilibrium Green function (NEGF) method, 80% spin injection efficiency (SIE) and 300% magnetoresistance ratio are predicted in multiple 2D barrier layers on the performance of MTJs. This is an effective mean to improve spin polarization in such junctions. There are also some works in the same direction. For example, Zhang *et al.*[138] investigated the vertical transport across $M/MoS_2/M$ (M = Co and Ni) MTJs with $MoS_2$ layer numbers N = 1, 3, and 5. Their results revealed that the thinner junctions are metallic because of the strong coupling between $MoS_2$ and the ferromagnets, and the junctions with thicker $MoS_2$ begin to show tunneling effects. A higher MR is achieved by increasing the number of interlayers. In their junctions, both positive (63.86%) and negative MR (-70.85%) can be obtained. A similar model based on their prediction was later developed experimentally. Galbiati *et al.*[139] reported the fabrication of NiFe/$MoS_2$/Co devices with mechanically exfoliated multilayer $MoS_2$ using an *in-situ* fabrication protocol that allows high-quality nonoxidized interfaces to be maintained between the ferromagnetic electrodes and the 2D layer. Their devices display a MR ratio up to 94%. Beyond interfaces and material quality, they suggested that spin-current depolarization could explain the limited MR. This points to a possible path towards the realization of larger spin signals in $MoS_2$-based MTJs.

Besides the number of layers of $MoS_2$, the length of scattering region can affect the MR ratio. By employing a three-band tight-binding model combined with the NEGF method, Jin *et al.*,[140] studied the spin-dependent electron transport in a zigzag monolayer $MoS_2$ with ferromagnetic electrodes. Their results reveal that the conductance shows a quantized oscillating phenomenon in the P configuration, while the conductance exhibits a zero platform in a large-energy region in the AP configuration. In addition, the length of the central part of the structure has a certain influence on the MR ratio. It is found that as the length of the middle region increased, the MR ratio decreased gradually. However, this prediction has not been approved by experiments yet. It hoped that the experimental reported would appear in the future.

Beyond non-magnetic $MoS_2$, 2D intrinsic magnetic materials play an important role in the spin polarization of MTJs. The reported molecular beam epitaxial growth of 2D magnetic materials for $Fe_3GeTe_2$, $VSe_2$, $MnSe_x$, and $Cr_2Ge_2Te_6$ opens new possibility for MR devices. They are magnetic conductors or insulators, which provide diverse application perspectives. For example, magnetic insulators are ideal for central tunneling layer in MTJs. And $CrI_3$ is a typical example of 2D magnetic insulating materials that have emerged in recent years. Atomically thin $CrI_3$ flakes were fabricated recently by mechanical exfoliation of bulk crystals onto oxidized silicon substrates.[124]

In $CrI_3$ flakes, intrinsic ferromagnetism and out-of-plane magnetization is observed. Interestingly, it is FM in the monolayer but becomes AFM in bilayer and back to FM in both trilayer and bulk.[141] And n-layer $CrI_3$ in the high-temperature phase exhibits inter-layer AFM coupling, which provides natural pinning layer for $CrI_3$. Yan *et al.*[142] studied the electron transport properties of $CrI_3$/BN/n-$CrI_3$ (n=1, 2, 3, 4) MTJs as shown in **Figs. 5 (a-d)**, and found the n=3 MTJ shows a fully polarized spin current with ~3600% TMR ratio when at the equilibrium state. More interestingly, the odd-even effect appears due to the difference of the number of pinning layers. The usage of different number of $CrI_3$ pinning layers greatly regulates the spin polarization.

As $CrI_3$ is a semiconductor which can serve as a *spin-filter tunnel barrier* when sandwiched between graphene electrodes, Song *et al*. performed an experimental work for improving spin polarization in MTJs by increasing the thickness of $CrI_3$ layers (**Figs. 5 (e-g)**).[143] The TMR of this spin-filter MTJs (sf-MTJs) can be drastically enhanced with the increase of the thickness of $CrI_3$ layers, which is corresponding with the former theoretical results.[142] When the thickness of $CrI_3$ increases to four layers, the TMR ratio can reach 19,000% at low temperature. The four-layer $CrI_3$ MTJ points to the potential for using layered antiferromagnets for engineering multiple magnetoresistance states in an individual multiple-spin-filter MTJ. The low $T_C$ of $CrI_3$ (around 50 K) limits its practical device application. It is urgent to find intrinsic 2D magnetic materials with high Curie or Neel temperature for room-temperature MTJ devices.

A similar theoretical investigation of a 2D spin filter and spin-filter MTJs consisting of atomically thin $Fe_3GeTe_2$ was also reported.[144] The models and the main data are shown in **Figs. 6 (a-d)**. By the DFT-NEGF method, the TMR effect is obtained in single/double-layer $Fe_3GeTe_2$−hBN−$Fe_3GeTe_2$ heterostructures. For heterostructures consisting of single- and double-layer $Fe_3GeTe_2$, the calculated MR ratio is 183% and 252% at zero bias, respectively. The $Fe_3GeTe_2$ MTJ in the P state shows a spin polarization of more than 75%.

Besides the use as a spin-filter barrier, the metallic nature of $Fe_3GeTe_2$ also enables itself to be used as magnetic electrodes in vdW MTJs to provide high spin polarization, which has advantages over insulating $CrI_3$ that is used as a spin-filter barrier only. On the one hand, a large magnetic field up to 1 T is required in $CrI_3$-based MTJs to switch the antiferromagnetic ground state to ferromagnetic.[145, 146] On the other hand, $CrI_3$-based MTJs are volatile, i.e., the magnetic field needs to be maintained to preserve the ferromagnetic order, while $Fe_3GeTe_2$-based MTJs are nonvolatile due to two stable P and AP magnetization configurations still appear under the absence of applied field. For this 2D metal with room-temperature ferromagnetism,[147] Li *et al.*[148] studied the spin-dependent electron transport across vdW MTJs consisted of a graphene or h-BN spacer layer with $Fe_3GeTe_2$ ferromagnetic electrodes (see **Figs. 6 (e-f)**). The authors found the resistance changes by thousands of percent from the P to AP state for both (graphene and h-BN spacers). The two different electronic structures of conducting channels in $Fe_3GeTe_2$ arouse a remarkable TMR effect. The authors further argued that the strain and interfacial distance are two main factors that can influence this giant TMR ratio. Also using $Fe_3GeTe_2$ as ferromagnetic electrodes, Zhang *et al.*[149] designed $Fe_3GeTe_2$|InSe|$Fe_3GeTe_2$ MTJs and found TMR reached ~700% in this kind of MTJs. By analyzing both complex band structure of the barrier and band structure of the electrode, the origin of the considerable TMR is disclosed, as shown in **Figs. 6 (h-i)**.

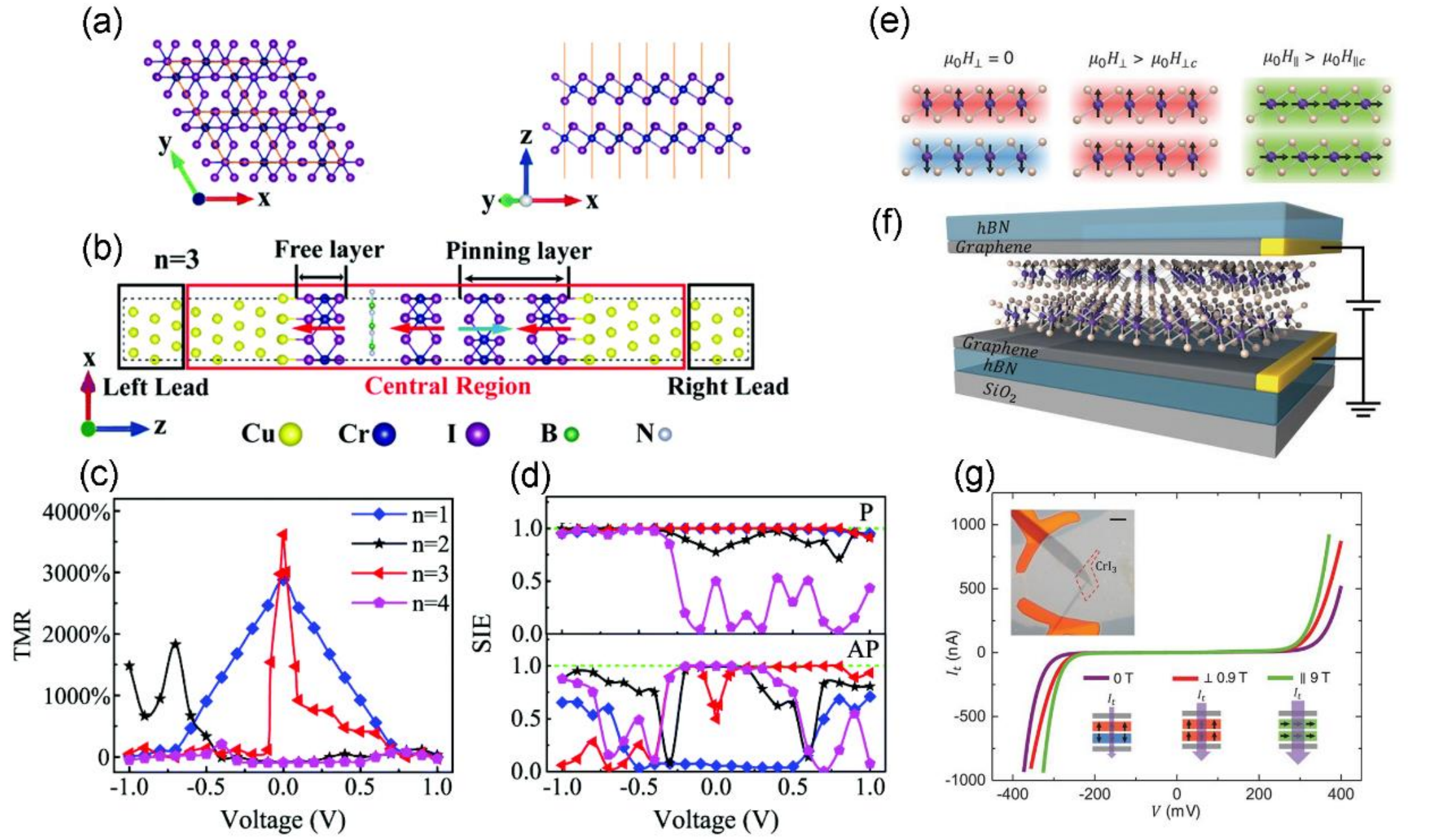

**Fig. 5** (a) Top and side views of bilayer $CrI_3$ in the high-temperature phase. (b) The model structures of Cu/$CrI_3$/BN/n-$CrI_3$/Cu (n=3, and 2 of them are the AFM pinning layer). (c) TMR versus bias voltage of Cu/$CrI_3$/BN/n-$CrI_3$/Cu, (d) SIE versus bias voltage of Cu/$CrI_3$/BN/n-$CrI_3$/Cu in P and AP configurations. *(a-d) Reproduced with permission from Yan et al., Phys. Chem. Chem. Phys.* ***22****, 26 (2020). Copyright 2020 Royal Society of Chemistry.* (e) Schematic of magnetic states in bilayer $CrI_3$. The left panel: layered-antiferromagnetic state, which suppresses the tunneling current at zero magnetic field. The middle and right panel: fully spin-polarized states with out-of-plane and in-plane magnetizations, which do not suppress it. (f) Schematic of 2D sf-MTJ. (g) Tunneling current of a bilayer $CrI_3$ sf-MTJ at selected magnetic fields. The top inset: optical microscope image of the device with scale bar of 5 μm. The red dashed line shows the position of the bilayer $CrI_3$. The bottom panel: schematic of the magnetic configuration for each $I_t$-V curve. *(e-g) Reproduced with permission from Song et al., Science* ***360****, 1214 (2018). Copyright 2018 The American Association for the Advancement of Science.*

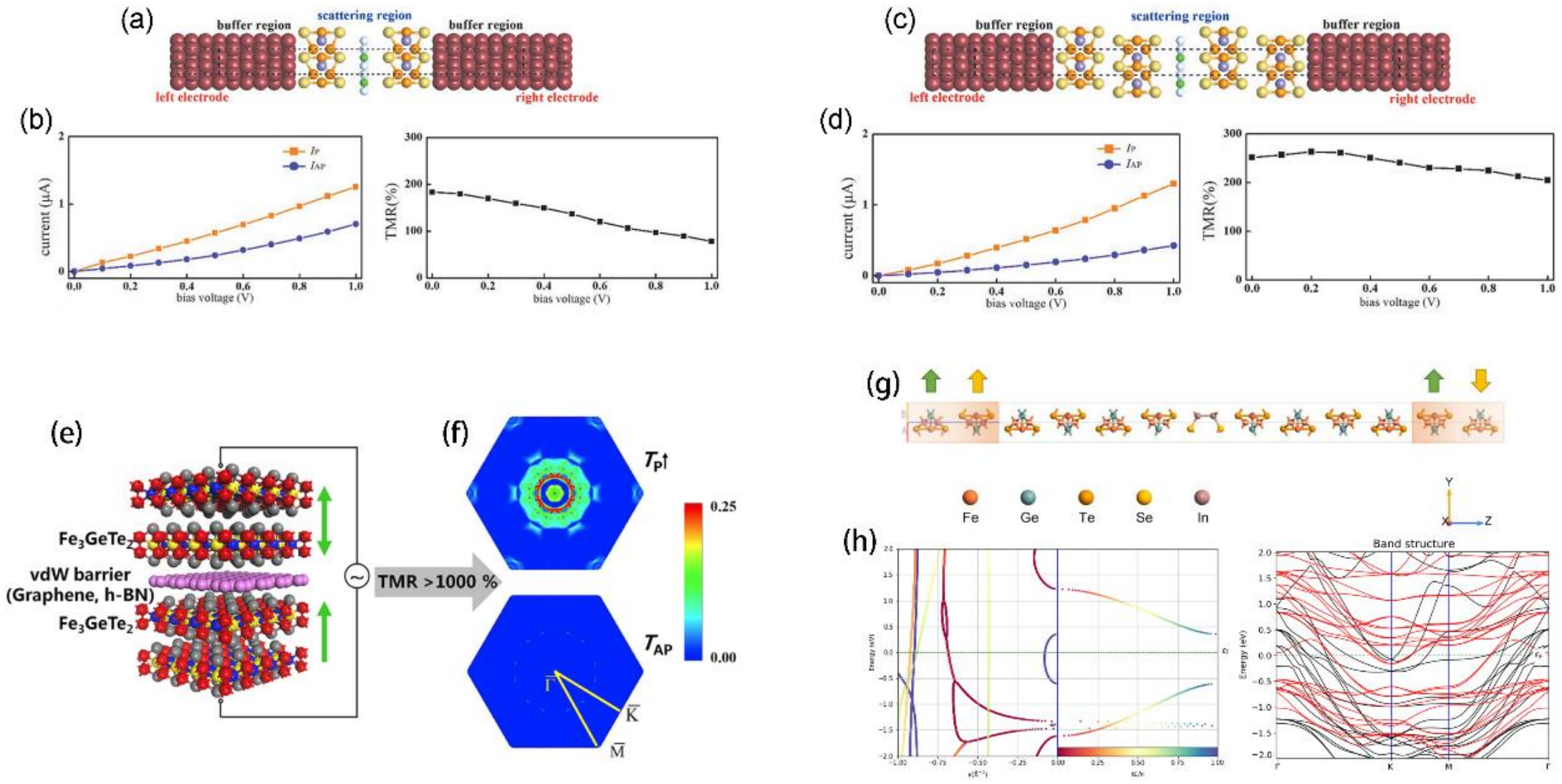

**Fig. 6** (a) The model of single−layer $Fe_3GeTe_2$ sandwiched between two Cu electrodes. (b) Spin-involved I-V curves and TMR of current varying with bias voltage of the model corresponding to (a). (c) The model of double−layer $Fe_3GeTe_2$ sandwiched between two Cu electrodes. (d) Spin-involved I-V curves and TMR of current varying with bias voltage of the model corresponding to (c). *(a-d) Reproduced with permission from Lin et al., Adv. Electron. Mater.* ***6****, 1900968 (2020). Copyright 2020 WILEY‐VCH.* (e) The magnetic junction structures of $Fe_3GeTe_2$/h-BN/$Fe_3GeTe_2$. (f) Electron tunneling probability in MTJs distribution in 2D brillouin region. *(e-f) Reproduced with permission from Li et al., Nano Lett.* ***19****, 5133 (2019). Copyright 2019 American Chemical Society.* (g)The structure of $Fe_3GeTe_2$|InSe|$Fe_3GeTe_2$ MTJ. (h) Complex band structure for the InSe part (left panel) and band structure of $Fe_3GeTe_2$ (right panel). *(g-h) Reproduced with permission from Zhang et al. J. Phys. Chem. C* ***124****, 27429 (2020). Copyright 2020 American Chemical Society.*

In the studies of the influence of the number of layers on the MTJ performance, it turns out that the number of layers of the central barrier layer has a great influence on the value of TMR (see more details in **Table 1)**.

**Table 1** Central and electrodes materials, defined TMR and TMR ratio with the function of the number of spin-filter tunnel barrier layers.

| Central materials | Electrode materials | TMR_max | Defined TMR | Remarks | Reference |
|---|---|---|---|---|---|
| $MnSe_2$/1-layer h-BN/$MnSe_2$ | Ir | 222.12% | $TMR = \frac{G_P - G_{AP}}{G_{AP}}$ | G is conductance | 150 |
| | Ru | 419.94% | | | |
| $MnSe_2$/2-layer h-BN/$MnSe_2$ | | 725.07% | | | |
| $VSe_2$/1-layer h-BN/$VSe_2$ | Ir | 254.48% | | | |
| | Ru | 183.55% | | | |
| $VSe_2$/2-layer h-BN/$VSe_2$ | | 199.15% | | | |
| 1-Layer $MoS_2$ | Co(fcc) | 63.86% | $TMR = \frac{G_P - G_{AP}}{G_{AP}}$ | G is conductance | 138 |
| | Co(hcp) | 33.40% | | | |
| | Ni(fcc) | 5.30% | | | |
| 2-Layer $MoS_2$ | Co(fcc) | 58.80% | | | |
| | Co(hcp) | 55.60% | | | |
| | Ni(fcc) | −13.79% | | | |
| 3-Layer $MoS_2$ | Co(fcc) | −70.85% | | | |
| | Co(hcp) | 55.91% | | | |
| | Ni (fcc) | 55.91% | | | |
| 1-Layer $MoS_2$ | $Fe_3Si$ (100) | 109.44% | $TMR = \frac{G_P - G_{AP}}{G_{AP}}$ | G is conductance | 151 |
| 3-Layer $MoS_2$ | | 306.95% | | | |
| 5-Layer $MoS_2$ | | 278.87% | | | |
| 7-Layer $MoS_2$ | | 154.56% | | | |
| 9-Layer $MoS_2$ | | 121.63% | | | |
| $CrI_3$/h-BN/1-layer $CrI_3$ | Cu (111) | 2900% | $TMR = \frac{I_P - I_{AP}}{I_{AP}}$ | I is current | 142 |
| $CrI_3$/h-BN/2-layer $CrI_3$ | | 1800% | | | |
| $CrI_3$/h-BN/3-layer $CrI_3$ | | 3600% | | | |
| $CrI_3$/h-BN/4-layer $CrI_3$ | | ~0% | | | |
| 1-Layer $Fe_3GeTe_2$/h-BN/1-layer $Fe_3GeTe_2$ | Cu | 183% | $TMR = \frac{1/I_{AP} - 1/I_P}{1/I_P}$ | I is current | 144 |
| 2-Layer $Fe_3GeTe_2$/h-BN/2-layer $Fe_3GeTe_2$ | | 289% | | | |
| 1-Layer $Fe_3GeTe_2$/h- | Cu | 78% | | P = $(D_{up}-D_{down})/(D_{up}+D_{down}$ | |
| BN/1-layer $Fe_3GeTe_2$ | | | $TMR_{Julliere} = \frac{1/I_{AP} - 1/I_P}{1/I_P} = \frac{2P^2}{1-P^2}$ | ) is the spin polarization of electrode; $D_{up}$ and $D_{down}$ are the spin−up and spin−down density of states in the Fermi level of electrode | |
| 2-Layer $Fe_3GeTe_2$/h-BN/2-layer $Fe_3GeTe_2$ | | 520% | | | |
| 1-Layer $Fe_3GeTe_2$/h-BN/1-layer $Fe_3GeTe_2$ | 2-Layer $Fe_3GeTe_2$ | 160% | $TMR_{scattering} = \frac{1/I_{AP} - 1/I_P}{1/I_P} = \frac{(D_{up} - D_{down})^2}{2D_{up}D_{down}}$ | T is transmission | |
| 2-Layer $Fe_3GeTe_2$/h-BN/2-layer $Fe_3GeTe_2$ | | 215% | | | |
| h-BN/2-layer $CrI_3$/$TaSe_2$/h-BN | Graphite | 240% | $TMR = \frac{R_{AP} - R_P}{R_P}$ | R is resistance | 19 |
| 2-layer $CrI_3$/$TaSe_2$ | | 40% | | | |

## 4.2 Targeting effective spin-injection

Note that in conventional MTJs, the electrodes usually adopt a kind of metal material, while the central scattering region is composed of an insulator material. Thus, the electrodes and the central scattering region are usually composed of different materials. The contact between them plays a significant role in determining the transport properties.[152] Unfortunately, due to lattice mismatch and conductivity mismatch, the contact between metallic electrodes and the insulating barrier is usually poor, resulting in *low spin injection*, which weakens the device performance.[153, 154] If these problems are not solved well, even in the off state, electrons from one side of the electrode have possibility of tunneling through the central scattering region to the other side of the electrode, resulting in the leakage of current.[155] Accordingly, it is difficult to fabricate high-performance devices based on the present configuration in experiments. As such, resolving the contact problem and/or spin injection into the central region is important for developing high performance MTJs. On the other hand, the parcel on the BN would effectively reduce the spin mismatch problem. In this section, we will review the recently reported 2D MTJs which target at improving spin injection.

To resolve lattice mismatch between electrodes and the barrier layer for high spin injection, the most directly mean is to adopt the same material. One 2D material can achieve both metal and insulator with different phase, for example $MoS_2$ in H phase and T phase. By using this mean, Zhou *et al.* [156] constructed a kind of graphene-based MTJ as shown in **Figs. 7(a-b)**. They systematically studied the transport properties of the zigzag graphene nanoribbon (ZGNR) using DFT-NEGF method. Remarkably, a 100% SIE and a giant TMR of up to $10^7$ is predicted in their designed graphene-based MTJ device, which shows much better performance than that of traditional 3D MTJs.

Besides choosing the same material, selecting two materials from the same family can also minimize the mismatch problem. Inspired by the experimental synthesis of the magnetic layered crystal of $Mn_2GaC$, its 2D counterpart of the half-metallic $Mn_2CF_2$ MXene layer can serve as the magnetic electrode for MTJs. $Ti_2CO_2$ MXene can be chosen as the tunneling barrier, which is one of the most studied MXenes in both experiments and theories.[157-159] Balcı *et al.*[160] designed an MXene-based MTJ, as shown in **Figs. 7 (c-e)**. The highlight of their work is the electrodes and barrier layer materials they chose are from the same family, which avoiding the lattice mismatch problem. And also, the band gap of $Ti_2CO_2$ barrier layer is almost the same as the half-metallic gap of $Mn_2CF_2$ electrodes. Both the barrier and the electrodes have a common carbon layer that contributes the most to the transmission. Based on above, the proposed MTJ is match both in structure and electronic structure. The proposed MTJ also exhibits TMR with a peak value up to $10^6$ and average values above $10^3$ within the bias of ±1 V. And then, Ünal Özden Akkuş *et al.*[161] further studied a similar work that investigated characteristics of a $Ti_2CT_2$ (T=O or F) MXene-based device which consists of semiconducting $Ti_2CO_2$ and $Ti_2CF_2$ metallic electrodes. Their DFT-NEGF transport calculations suggest that the device, made of a $Ti_2CT_2CO_2$ semiconductor and two $Ti_2CF_2$ metallic electrodes, shows field-effect transistor characteristics when the semiconducting part is longer than about 6 nm. It is found that devices with larger tunneling barrier width should have a much better response to the gate voltage.

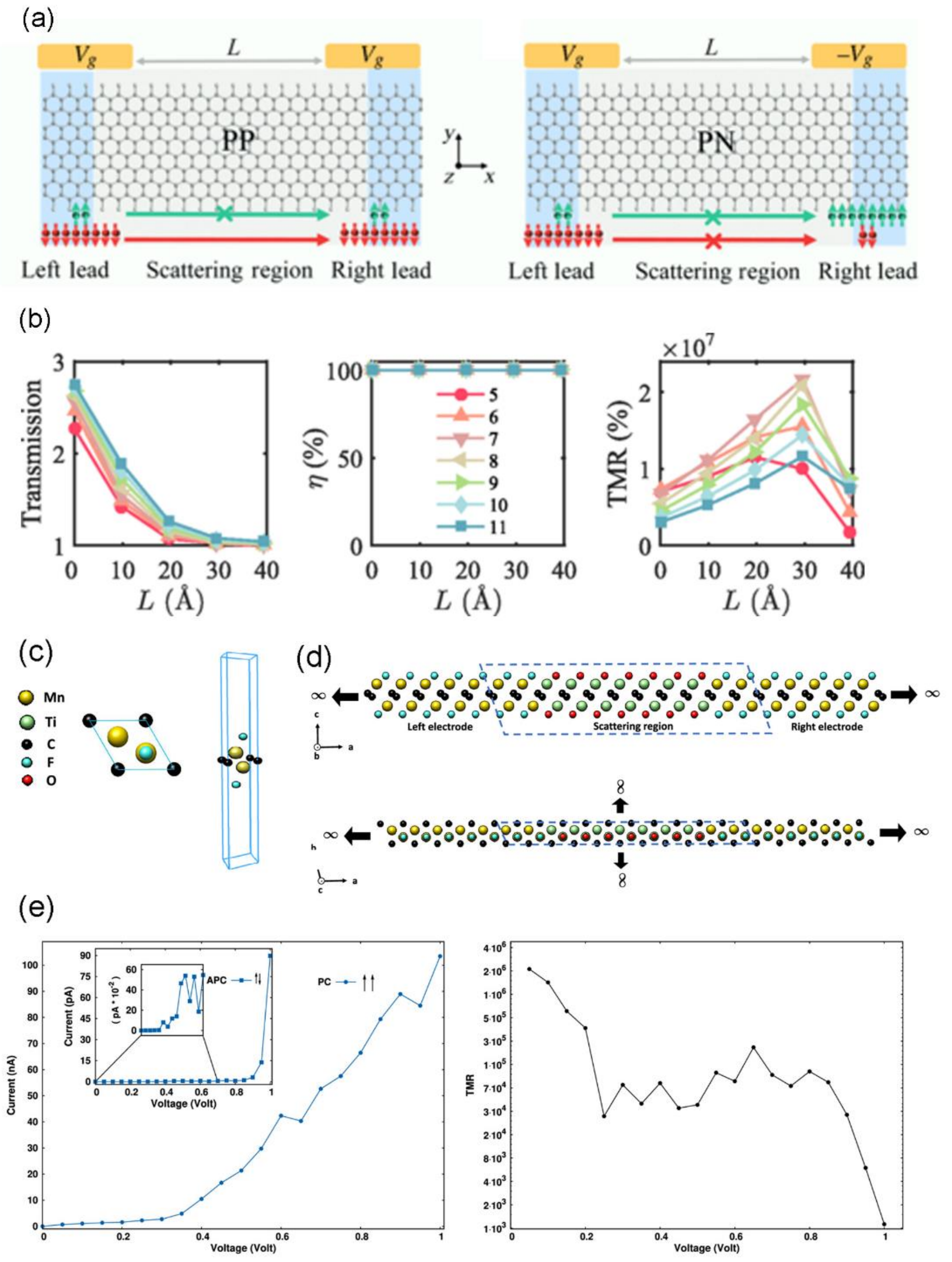

**Fig. 7** (a) The structures of the device where in-plane gate voltages are applied to the electrode regions in the same direction (PP) and opposite directions (PN). The length of the central scattering region without no gate voltage is applied, is denoted by L. (b) Spin-down transmission coefficient (left panel), spin injection efficiency (middle panel), and TMR (right panel) as a function of L for the PP configuration. *Reproduced with permission from Zhou et al. Phys. Rev. Appl.* ***13****, 044006 (2020). Copyright 2020 American Physical Society.* (c) Top view and side view of $Mn_2CF_2$ unit cell. (d) The top and side views of an MXene-based MTJ ($Mn_2CF_2/Ti_2CO_2/Mn_2CF_2$). (e) I-V curves and TMR of the MTJ in P and AP configurations. *Reproduced with permission from Balcı et al. ACS Appl. Mater. Interfaces* ***11****, 3609 (2018). Copyright 2019 American Chemical Society.*

Another effective way to solve the conductivity mismatch problem and improve spin injection in MTJs is the introduction of h-BN, as discussed in **Sec. 2**. The insulating h-BN is proposed as an ideal covalent spacer for MTJs, which provides a higher MR ratio and stronger exchange coupling at the interface to remove the conductivity mismatch between the metal leads and the FM layer. Many works in both theories and experiments have proved that h-BN based MTJs have a good TMR performance. Such conclusion has great impact on the h-BN integration pathway. Begin with the theoretical works, Qiu *et al.,*[162] proposed an effective method to control the spin current in a vertical MTJ by combining the strong spin filtering effect of graphene/ferromagnet interface with the resonant tunneling effect of graphene/h-BN/graphene vdW heterostructure, in which Ni(111) is used as electrodes. Their theoretical results reveal that when the electronic spectra of spin electrons in two graphene layers are aligned, the spin resonance would appear which results in a negative differential resistance (NDR) effect. By studying the similar structure with a periodic density functional method in conjunction with Julliere's model, Sahoo *et al.*[163] constructed Ni/BN/graphene/BN/Ni and Ni/graphene/Ni tunnel devices. It is found that the former, the graphene/h-BN multi-tunnel junction, has a much higher TMR than the latter. The underlying mechanism was explained by Wu *et al.*[164] that the minority-spin-transport channel of graphene can be strongly suppressed by the insulating h-BN barrier, which overcomes the spin-conductance mismatch between Ni and graphene, resulting in a high spin-injection efficiency. Another example of using h-BN as the tunnel barrier, is $MnSe_2$/h-BN/$MnSe_2$.[150] The schematic diagram of the proposed structure is shown in **Fig. 8**. The monolayer T-$MnSe_2$ is selected as the ferromagnetic layer,[165] and Ir and Ru are employed as metal electrodes, which have a smaller lattice mismatch with $XSe_2$ compared with conventionally well-used electrode metals (such as Au, Ag and Cu). It is found that such vertically vdW MTJ has a large TMR of 725%.[150] This is due to a large transmission in majority channel in the P magnetic configuration, while it is suppressed in the AP magnetic configuration, as shown in **Figs. 8 (b)** and **(e)**. The authors take two approaches for improving TMR: one is choosing suitable electrodes and another is finitely increasing the number of layers of the h-BN barrier. The former minimizes the lattice mismatch and the latter involving BN solves the spin mismatch problem.

As discussed above, introducing extra 2D BN layer has advantages in improving TMR than 3D tunneling barriers like $Al_2O_3$, which has been supported in experiments successfully[106]. And many efforts have been done in the actual manufacture of BN-based MTJs to improve TMR further. For example, a TMR ratio around 0.3-0.5% can be obtained by performing wet transfer on ferromagnet,[166] while h-BN is exfoliated on perforated membranes the TMR ratio is 1%,[167] and the TMR can research up to 6% grown by large area CVD on Fe directly.[168]

Besides using calculations to predict the conclusions like above discussions, using calculations to explain experiments is also important and necessary due to calculations can give some physical insights. Liking the research on different ferromagnets to develop h-BN MTJs, it is particularly important to using calculations to understand the interfacial effect on the MR performance of MTJs. For example, Piquemal-Banci *et al.*[169] fabricated two h-BN-based MTJs with different FM electrodes, Co/h-BN/Co and Co /h-BN/Fe MTJs. In these two MTJs, h-BN is grown directly by CVD on pre-patterned Co and Fe stripes. The TMR ratio in these two MTJs are 12% and 50%, respectively. By using calculations method, the expected strong dependence of the h-BN electronic properties on the coupling with the FM electrode is further investigated. This calculations part gives an explanation of how h-BN improves TMR in their experimental conclusion.

When h-BN acts as the tunneling barrier, the FM layers can be various beyond Fe and Co. Like $CrI_3$ which is introduced above, its cousin $CrBr_3$ monolayer also shows FM character.[141] And $CrX_3$ heterojunctions have been reported having large MR recently.[143] Inspired by this, Pan *et al.*[170] systematically investigated the structural, magnetic and spin-transport properties of $CrX_3$/h-BN/$CrX_3$ (X = Br, I) MTJs. The barrier layer is h-BN and $CrX_3$ are used as the ferromagnetic layer. Metal Au, Ag, Al, and Pt are chosen as electrodes. Taking the Ag$CrBr_3$/h-BN/$CrBr_3$/Ag MTJ (**Fig. 8 (f)**) as an example, the large TMR effect can be up to 1,565% in this series of MTJs. The diagram of the band alignment and the k-resolved transmission spectra of MTJ are shown in **Figs. 8 (g-k)**.

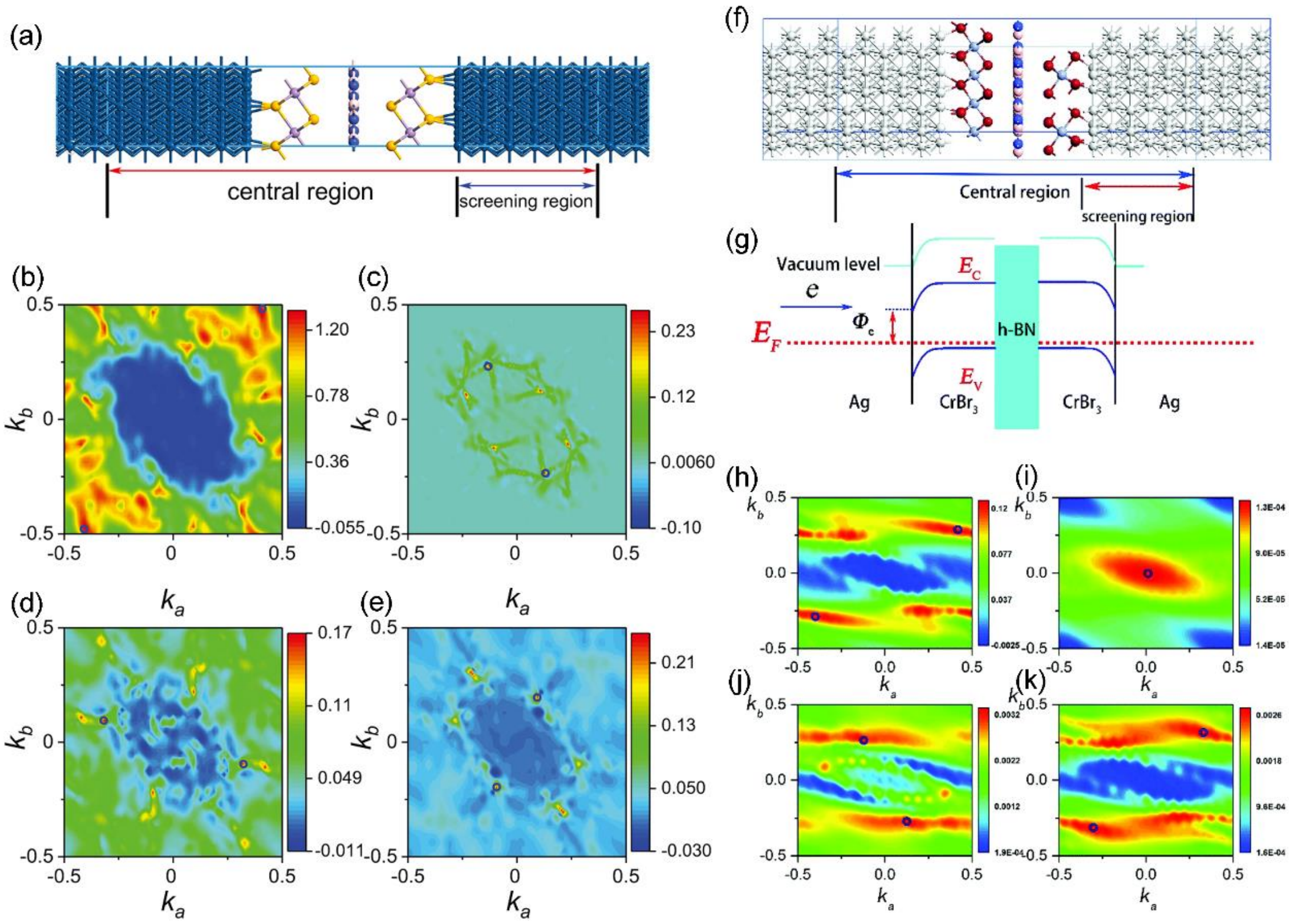

**Fig. 8** (a) The structure of Ru/$MnSe_2$/h-BN/$MnSe_2$/Ru MTJs. (b-e) $k_{||}=(k_a,\ k_b)$

dependent transmission spectra of MTJs based on $MnSe_2$ for majority-spin and minority-spin states in P and AP configurations, respectively. *(a-e) Reproduced with permission from Pan et al., Chin. Phys. B **28**, 107504 (2019). Copyright 2019 CPB.* (f) The structure of $CrBr_3$/h-BN/$CrBr_3$ MTJs. White colored atoms on the left and right sides are silver electrodes. (g) The diagram of the band alignment of this MTJ. (h-k) The k-resolved transmission spectra of MTJ for majority- and minority-spin states in P and AP configurations. *(f-k) Reproduced with permission from Pan et al., Nanoscale **10**, 22196 (2018). Copyright 2018 Royal Society of Chemistry.*

### 4.3 Targeting spin-manipulation

Manipulating spin is another powerful mean not only to improve TMR, but also to realize different functionalities of spintronic devices. There are many ways for spin manipulation, such as adding other vdW materials to induce local phase transition, introducing SOT, making use of interlayer interaction, doping FM, and so on. In this section, we mainly review recent works which target spin manipulation of MTJs.

For the aspect of spin manipulation through phase transition induced by an adjacent material, Begunovich *et al.* [171] proposed an ultrathin MTJs based on vanadium ditelluride monolayers with graphene as a tunnel barrier. Both trigonal prismatic (H-phase) and octahedral (T-phase) $VTe_2$ were considered in their study. The authors found that the introduction of graphene makes the electronic characteristic of 2D T-$VTe_2$ changeable from the metal to half-metal phase, making T-$VTe_2$ a promising candidate for MTJ applications. Although several possible structures are considered, the one which follows the framework of Julliere model shows the highest TMR ratio up to 220%.

Inserting other vdW "heavy" materials may introduce SOT-driven operations. Combining NEGF with *noncollinear* density functional theory (ncDFT) methods, Dolui *et al.*[19] constructed bilayer-$CrI_3$/monolayer-$TaSe_2$ vdW lateral heterostructure as shown in **Fig. 9 (a)**, in which bulk non-magnetic metal electrodes are required in practice for generating in-plane charge current. They found that the AFM–FM nonequilibrium phase transition can be induced by the SOT in this MTJs where the unpolarized charge current is injected parallel to the interface. The 1H phase monolayer of metallic $TaSe_2$ is chosen because of a small lattice mismatch (0.1%) to $CrI_3$ and inversion asymmetry. By introducing another heavy metal of $WS_2$, Zollner *et al.*[36] designed $Cr_2Ge_2Te_6$/graphene/$WS_2$ vdW MTJs (see **Fig. 9 (b)**), where SOC, valley-Zeeman and Rashba splitting, and exchange coupling can be obtained.

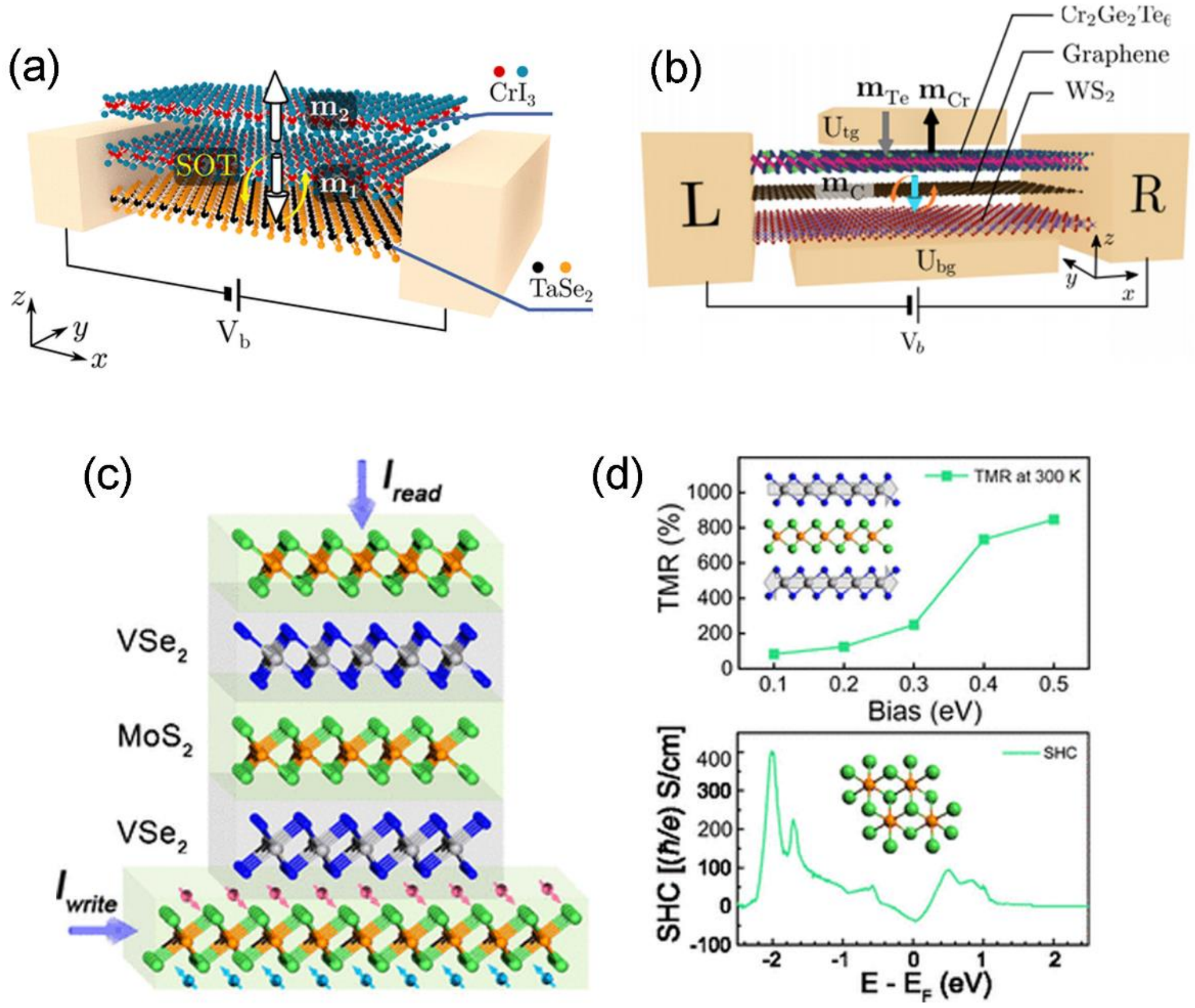

**Fig. 9** (a) The geometric structure of the $CrI_3$/$TaSe_2$ SOT vdW MTJ. *(a) Reproduced with permission from Dolui et al., Nano Lett.* ***20****, 2288 (2020). Copyright 2020 American Chemical Society.* (b) The structure of the $Cr_2Ge_2Te_6$/graphene/$WS_2$ vdW MTJ. *(b) Reproduced with permission from Zollner et al., Phys. Rev. Res* ***2****, 4 (2020). Copyright 2020 American Physical Society.* (c) The structure of the $VSe_2$/$MoS_2$ SOT vdW MTJ. (d) TMR and SHC of $VSe_2$/$MoS_2$ SOT vdW MTJ. *(c-d) Reproduced with permission from Zhou et al., ACS Appl. Mater. Interfaces* ***11****, 17647 (2019). Copyright 2019 American Chemical Society.* The key feature of this type of SOT-MTJs is that the write charge current horizontally flows a heavy non-magnetic 2D material, while the read spin current vertically flows a 2D vdW MTJ.

As for enhancing TMR by making use of interlayer interaction, many efforts are also done. To understand the inside mechanism better, calculations method is needed. For example, Heath *et al.*[172] shed important insights from an atomistic viewpoint on the underlying mechanism governing the spin transport in graphene/$CrI_3$ spin-filter tunneling junctions by a combined first-principles and quantum ballistic transport calculation, as shown in **Fig. 10**. The calculated electronic structures reveal that tunneling is the dominant transport mechanism in these heterostructures. The tunneling effect boosts differentiate intermediate metamagnetic states presenting in the switching process. This is manifested in an increase in TMR for energy above the Fermi level due to enhancement of Bloch states near the edge of the conduction band of $CrI_3$.

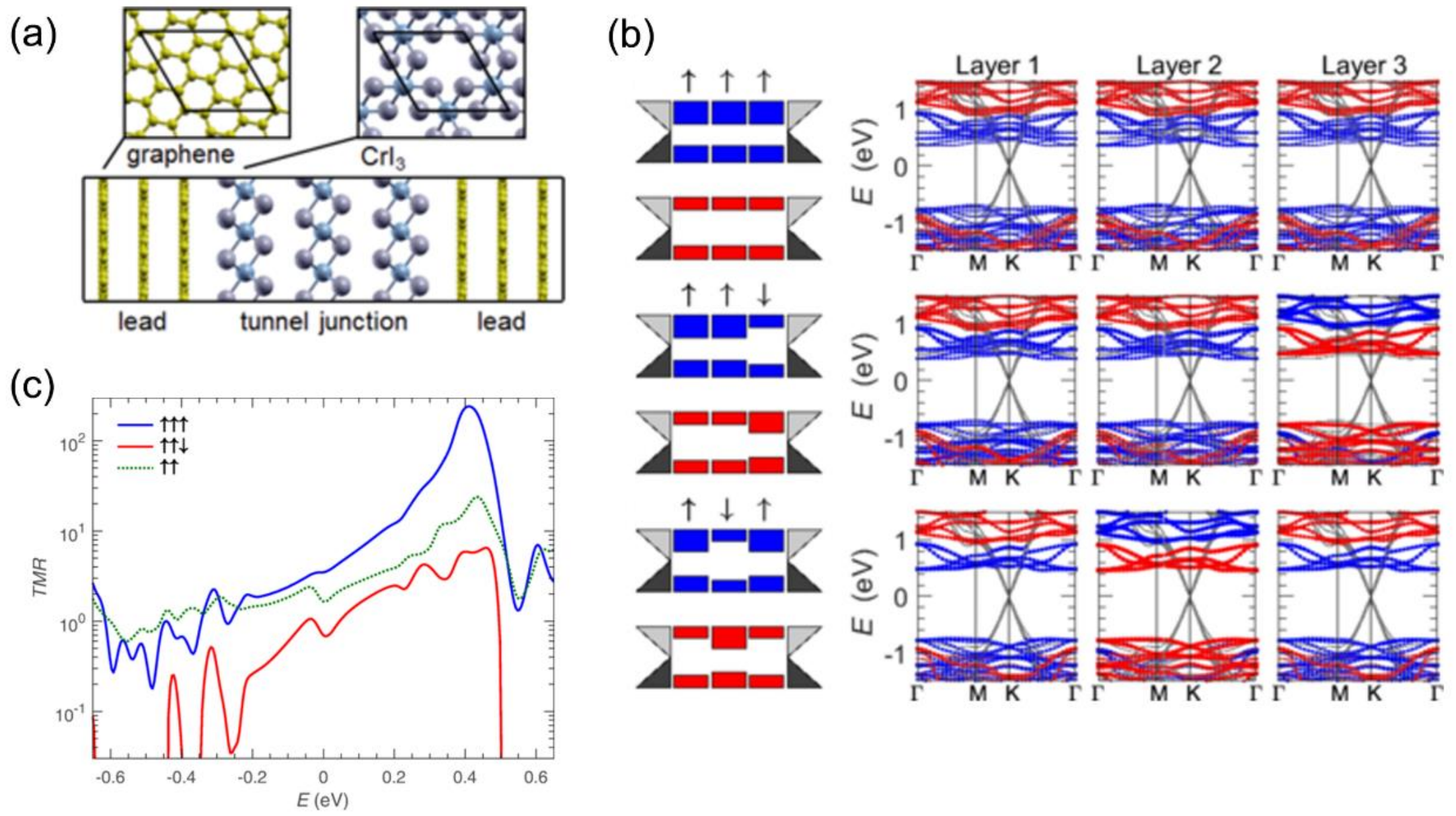

**Fig. 10** Atomic and electronic structures of graphene/$CrI_3$/graphene MTJs. (a) Atomic structures of graphene/$CrI_3$/graphene MTJs. (b) Band diagrams and corresponding band structures in trilayer graphene/trilayer $CrI_3$/trilayer graphene junctions for various states. (c) TMR as a function of Fermi level for both trilayer (TMR↑↑↑ and TMR↑↑↓) and bilayer (TMR↑↑) systems. *Reproduced with permission from Heath et al., Phys. Rev. B* ***101****, 195439 (2020). Copyright 2020 American Physical Society.*

In order to manipulate spin in non-magnetic 2D materials, one can dope them by charge transfer from FM metals or proximity-induced spin splitting in themselves. For example, Asshoff *et al*.[167] fabricated vertical graphene-based devices where ultimately clean graphene–FM interfaces were obtained by depositing the FM metals (FM = Co and FM' = $Ni_{0.8}Fe_{0.2}$ alloy) on the two sides of a suspended graphene membrane, thereby preventing oxidation, minimizing the number of fabrication steps and limiting the exposure of the devices to solvents during preparation. Such kind of treatment improves the performance of MTJs.

Applying a *finite external bias voltage* has been proved to be an effective method to manipulate spin transport. For example, Chen *et al*.[173] theoretically investigated the nonequilibrium spin injection and spin-polarized transport in monolayer black phosphorus (MBP) with ferromagnetic Ni contacts. The top and side views of their model are shown in **Fig. 11**. In this study, they explored the SIE, TMR ratio, spin-polarized currents, charge currents and transmission coefficients as a function of bias voltage. Furthermore, they studied two different contact structures where MBP is contacted by Ni(111) and Ni(100). Both structures are predicted to have great spin-polarized transport performance.[173] The Ni(100)/MBP/Ni(100) MTJ has the superior properties of the SIE (~60%) and TMR ratio (40%), which maintains almost a constant value against the bias voltage.

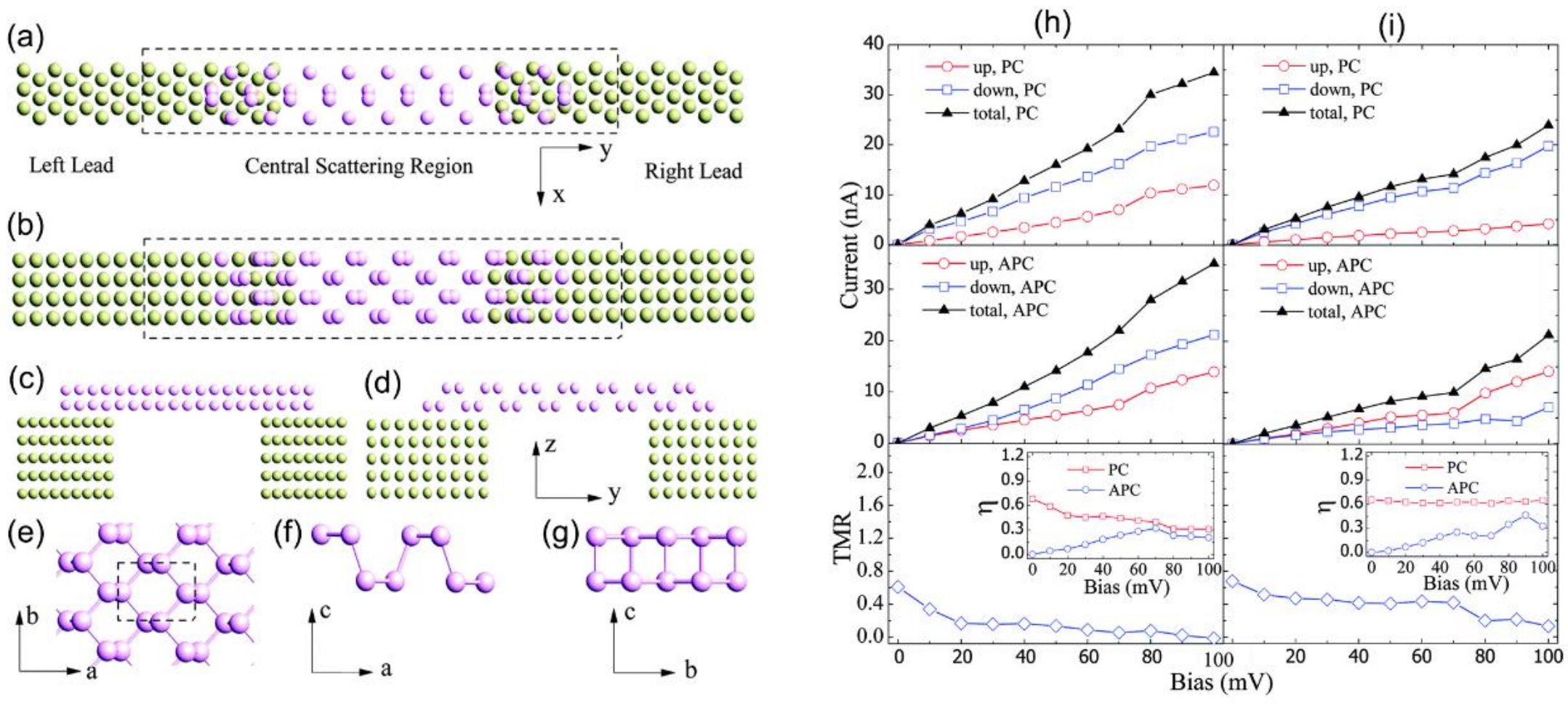

**Fig. 11** (a) The structure of Ni(111)/MBP/Ni(111) MTJ in the top view. (b) The structure of Ni(100)/MBP/Ni(100) MTJ in the top view. (c-d) The side view of Ni(111)/MBP/Ni(111) MTJ and Ni(100)/MBP/Ni(100) MTJ respectively. (e-g) The top and side view of MBP. Ni and P atoms are yellow and pink, respectively. (h-i) I-V curves, TMR and SIE of Ni(111)/MBP/Ni(111) and Ni(100)/MBP/Ni(100) MTJ, respectively. *Reproduced with permission from Chen et al., Phys. Chem. Chem. Phys.* ***18****, 1601 (2016). Copyright 2016 Royal Society of Chemistry.*

Hydrostatic pressure can be used for continuous control of interlayer coupling by interlayer spacing in vdW crystals, and then tuning the spin interaction and transport. For example, experimentally, Song *et al.*[174] demonstrated the changes of magnetic order by pressure in 2D magnet $CrI_3$. The MTJ structure is composed of bilayer/trilayer $CrI_3$ sandwiched by top and bottom multilayer graphene contacts, and h-BN encapsulates the whole MTJ in order to avoid sample degradation. **Figure 12** shows the structure of a bilayer $CrI_3$ MTJ. It is found that the interlayer magnetic coupling can be doubled by a hydrostatic pressure.[174]

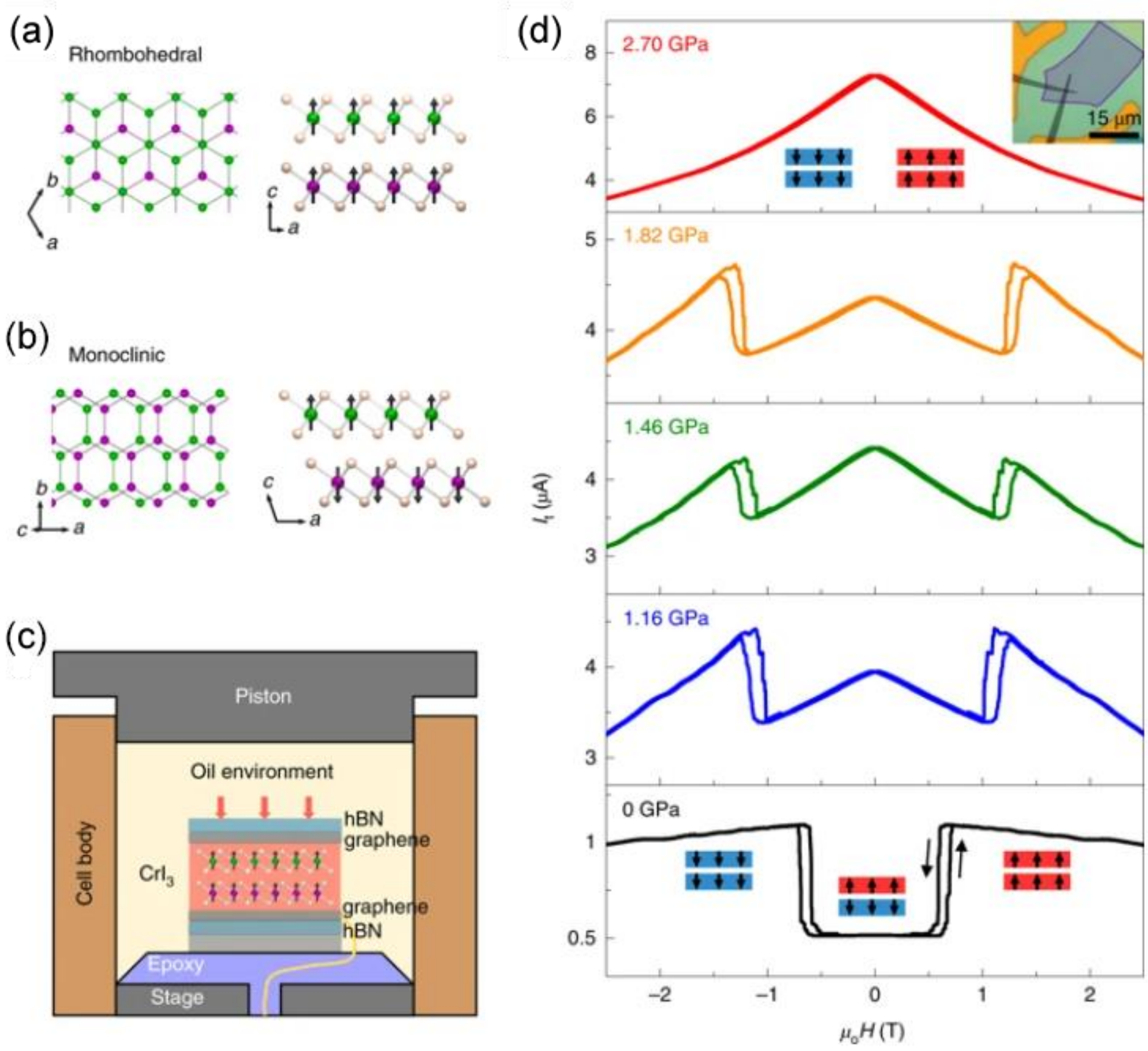

**Fig. 12** Stacking-order determined 2D magnetism and tunneling measurements of bilayer $CrI_3$ under pressure. *Reproduced with permission from Song et al., Nature materials, **18**, 12 (2019). Copyright 2019 Springer Nature.*

### 4.4 Targeting stability

Besides the target high-performance of MTJs, one should consider the structural and magnetic stabilities in the practical application. These practical problems include whether the MTJs can operate at room temperature, and whether the 2D materials used in MTJs can be successful synthesized. Target at these problems, many efforts have been made in 2D-materials-based MTJs. In this section, we would review the recently work around the efforts which is target at the stability problems.

In order to achieve 2D-materials-based MTJs, the first required is the 2D materials are stable in room-temperature. Thus, thermodynamic stability, dynamic stability, and mechanical stability should be assessed. In practical efforts, inert 2D materials like BN are often used to wrap reactive 2D materials like black phosphorene.[175]

Room-temperature working devices also require magnetic stability. However, most of 2D FM materials discovered to date suffer from low $T_C$. As a result, the MTJs constructed by these materials only works at low temperatures, such as $CrI_3$-based MTJs. High temperature working MTJs can be achieved by using 2D FM materials which have been predicted/discovered to have high $T_C$. Recently, monolayer $VSe_2$ is reported to be a room-temperature ferromagnetic 2D material experimentally.[176] Thus, a $VSe_2/MoS_2$ vdW MTJ was theoretically designed by Zhou *et al*,[20] as shown in **Figs. 9 (c-d)**. They proposed a concept of SOT vdW MTJs, which can achieve both reading and writing functions at room-temperature. Their NEGF results show a TMR up to 846%. This proposed SOT vdW MTJs based on $VSe_2/MoS_2$ give 2D MTJs a new opportunity for many magnetic-field-free device applications, which can work in room-temperature. Later on, more $XSe_2$ (X= Mn, V) based MTJs [150] with 300 K working temperature are proposed. The search for room temperature FM materials is always demanding for 2D MTJs. Recently, Yang *et al*.[177] designed excellent ultrathin spin filters by using half-metal 2D $Cr_2NO_2$ (see **Fig. 13**), which has a $T_C$ of 566 K, based on first-principles calculations. The half-metal feature with 100% spin polarization of $Cr_2NO_2$ guarantee a giant TMR up to 6,000%.

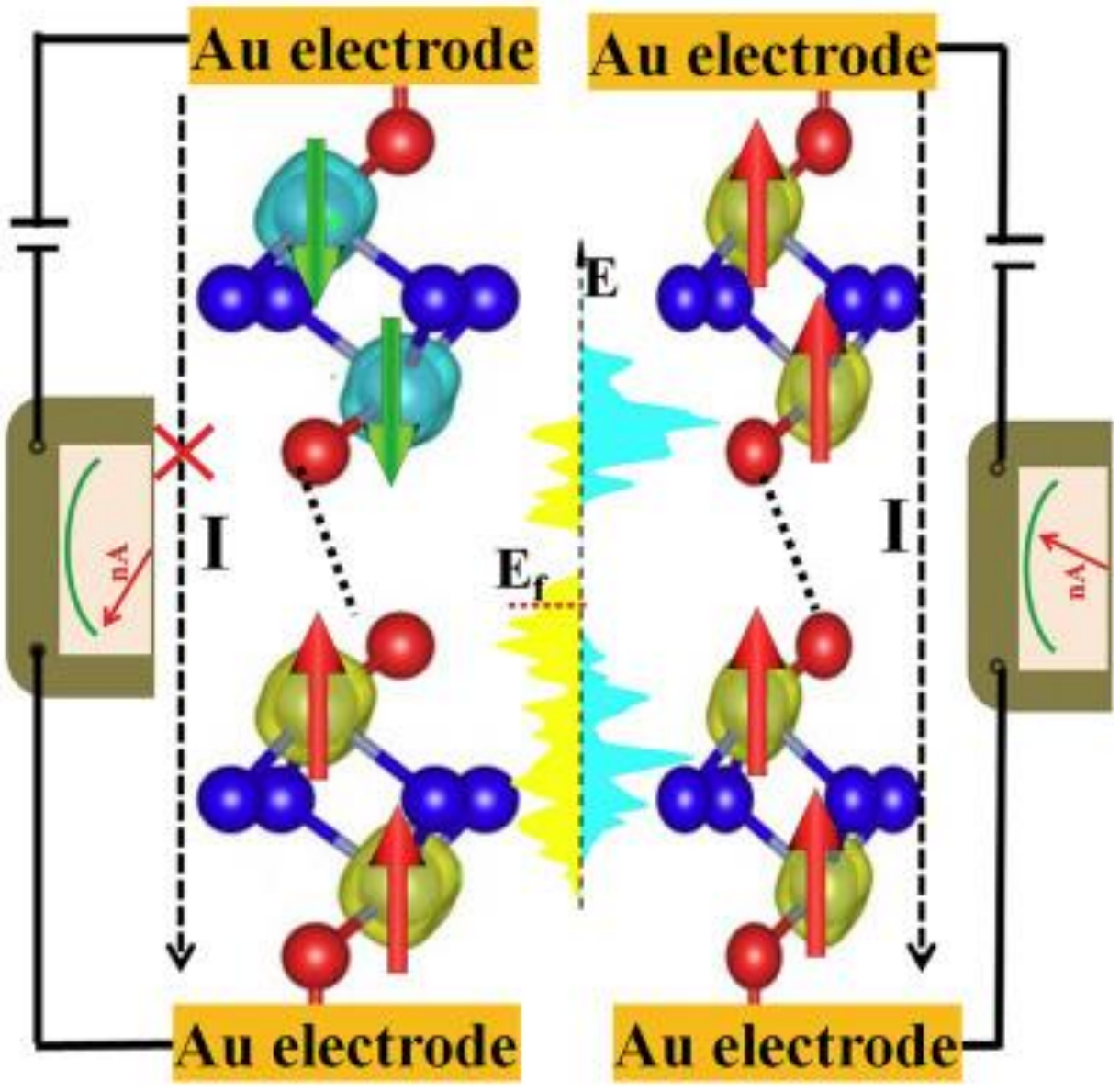

**Fig. 13** The structure of Au/bilay-$Cr_2NO_2$/Au. *Reproduced with permission from Yang et al., Matter **1**, 1304 (2019). Copyright 2019 Elsevier.*

Perpendicular MTJs with out-of-plane interfacial magnetization have many advantages, such as high thermal stability, infinite endurance, and fast with low-power switching. They are the base to construct advanced non-volatile memory device which can be built as non-von Neumann computing paradigms to overcome power bottleneck. Some perpendicular MTJs have been proposed theoretically,[178] and then was made in the experiment[41]. From the materials point of view, it is important and necessary to discover more 2D magnetic materials with out-of-plane magnetic anisotropy, which can be used as building blocks for perpendicular MTJs.

## 5. Summary and perspectives

Since the discovery of TMR in MTJs, the MTJ devices have shown a great and profound impact on spintronics applications, including the hard disk driver, MRAM, radio-frequency sensors, microwave generators and neuromorphic computing networks. In this review, we first start from four main issues of conventional MTJs, and then review the current progress of 2D-materials-based MTJs and how they address the problems of conventional MTJs followed by brief comments on the new scientific problems and technical challenges in 2D MTJs.

In the past decade, the emergence of 2D (magnetic) materials has brought fresh blood to the family of MTJs and its spintronics applications. The good performance of electron transport properties presented by graphene, h-BN, $MoS_2$, $CrI_3$, $Fe_3GeTe_2$ and their vdW heterostructures has led to the prediction and demonstration of perfect spin filtering and large TMR ratio, as elaborated in this review. A TMJ usually is a stack of two or more materials in terms of a FM-NM-FM configuration. The various ways of stacking of concerned 2D materials result in 2D MTJs with diverse structures, which are categorized as lateral and vertical vdW MTJs. The 2D materials can be the magnetic electrodes and/or the central insulating materials. Many factors have been reported to influence the performance of 2D MTJs, such as the thickness, strain, interface and the number of layers of 2D materials, which have been discussed in this review.

For the computational aspect, the combination of DFT with the NEGF method is one of the most common technique to study the MTJs' transport properties. The main challenge of this method is the over-estimated TMR ratio. It is because under the framework of DFT the complicated interfacial configuration, such as disorder, and the temperature effect cannot be described properly due to the demand in computational resources. Some advanced methods in calculations of transport properties of MTJs are needed. In 2008, Ke *et al.*[179] developed a nonequilibrium vertex correction (NVC) theory to handle the configurational average of random disorder at the density matrix level so that disorder effects to nonlinear and nonequilibrium quantum transport can be calculated from atomic first principles in a self-consistent and efficient manner. Recently, Starikov *et al.*[180] described a DFT-based two-terminal scattering formalism that includes SOC and spin noncollinearity. An implementation using tight-binding muffin-tin orbitals combined with extensive use of sparse matrix techniques allows a wide variety of inhomogeneous structures to be flexibly modelled with various types of disorder including temperature induced lattice and spin disorder. The low stability and $T_C$ of 2D magnetic materials, such as $CrI_3$, hamper the practical applications of MTJ devices.

The future prospect in 2D MTJs' development might be keeping pace closely with the emergence of novel stable 2D magnetic materials with room-temperature $T_c$. However, the conventional DFT method is not quick enough to discover more demanded 2D materials with high $T_c$ and out-of-plane magnetic anisotropy. Recently, high-throughput computations have been carried out to screen high-$T$c 2D materials. For example, Jiang *et al.*[181] investigated the electronic and magnetic properties of 22 monolayer 2D materials with layered bulk phases. They showed that monolayer structures of CrSI, CrSCl, and CrSeBr have notably high $T_C$ (>500 K) and favorable formation energies, making these ferromagnetic materials feasible for experimental synthesis. Besides discovering of new 2D magnetic materials using high-throughput calculations, theoretically rational design of novel systems such as mixed dimensional heterostructures and mixed dimensional vdW heterostructures certainly provide a stable geometrical and magnetic structure as well as give the additional spin injection and spin-state control tools through the spin–orbit combined with the Coulomb effect, improving the MTJ device performance.

In recent years, more and more researches combine experimental work with computational work. One should design a calculation model that can better reflect the actual situation of the experiment, instead of being too idealistic. The calculation should take full account of the actual experimental conditions as far as possible, and not be too unconstrained.

As for experimental aspect, the key research direction to develop device mainly focuses on integration engineering and the understanding of 2D ferromagnet growth processes. At present, the research for this aspect is still at an early stage though some progress has already been made. In fact, some integration parameters are not understood clearly. Various methods are applied to growth 2D materials. Many parameters such as 2D materials' quality, crystallinity, phase, surface chemical state, thickness and the interaction at the interface would have great impact on the performance of MTJ devices. In addition to focus on improving TMR ratio of 2D MTJs, the device lifetime, mechanical stress, thermal stability and switching current are also important and have research value in the experiment. Finally, with the implementation of large-scale growth of advanced 2D materials through chemical vapor deposition methods, it is indubitable to expect a real promise exists for a new generation of 2D MTJs on a large scale.

## Acknowledgements

The authors would like to acknowledge the support from the National Natural Science Foundation of China (Grant No. 51671114 and No. U1806219), MOE Singapore (MOE2019-T2-2-030, R-144-000-413-114, R-256-000-651-114 and R-265-000-691-

114). This work is also supported by the Special Funding in the Project of the Taishan Scholar Construction Engineering.

## Data availability

Data sharing is not applicable to this article as no new data were created or analyzed in this study.

## Notes and references

The authors declare no conflict of interest.

[1] D. Akinwande, N. Petrone, and J. Hone, Two-dimensional flexible nanoelectronics, Nature communications **5**, 1 (2014).

[2] W. Lu, and C. M. Lieber, in *Nanoscience And Technology: A Collection of Reviews from Nature Journals* (World Scientific, 2010), pp. 137.

[3] R. Westervelt, Graphene nanoelectronics, Science **320**, 324 (2008).

[4] R. Chau *et al.*, Integrated nanoelectronics for the future, Nature materials **6**, 810 (2007).

[5] L. Zhang *et al.*, Diverse transport behaviors in cyclo [18] carbon-based molecular devices, The Journal of Physical Chemistry Letters **11**, 2611 (2020).

[6] S. Sanvito, Molecular spintronics, Chem. Soc. Rev. **40**, 3336 (2011).

[7] J. Linder, and J. W. Robinson, Superconducting spintronics, Nature Physics **11**, 307 (2015).

[8] W. Han *et al.*, Graphene spintronics, Nature nanotechnology **9**, 794 (2014).

[9] R. Jansen, Silicon spintronics, Nature materials **11**, 400 (2012).

[10] D. E. Bruschi, A. R. Lee, and I. Fuentes, Time evolution techniques for detectors in relativistic quantum information, Journal of Physics A: Mathematical and Theoretical **46**, 165303 (2013).

[11] B. Collins, and I. Nechita, Random matrix techniques in quantum information theory, Journal of Mathematical Physics **57**, 015215 (2016).

[12] L. M. Vandersypen, and I. L. Chuang, NMR techniques for quantum control and computation, Rev. Mod. Phys. **76**, 1037 (2005).

[13] Y. M. Lee *et al.*, Giant tunnel magnetoresistance and high annealing stability in Co Fe B ∕ Mg O ∕ Co Fe B magnetic tunnel junctions with synthetic pinned layer, Appl. Phys. Lett. **89**, 042506 (2006).

[14] G. Miao *et al.*, Epitaxial growth of MgO and Fe ∕ Mg O ∕ Fe magnetic tunnel junctions on (100)-Si by molecular beam epitaxy, Appl. Phys. Lett. **93**, 142511 (2008).

[15] C. Duret, and S. Ueno, TMR: A new frontier for magnetic sensing, NTN technical review **80**, 64 (2012).

[16] S. Mao *et al.*, Commercial TMR heads for hard disk drives: characterization and extendibility at 300 gbit 2, IEEE Trans. Magn. **42**, 97 (2006).

[17] R. W. Dave *et al.*, MgO-based tunnel junction material for high-speed toggle magnetic random access memory, IEEE Trans. Magn. **42**, 1935 (2006).
[18] D. C. Ralph, and M. D. Stiles, Spin transfer torques, J. Magn. Magn. Mater. **320**, 1190 (2008).
[19] K. Dolui *et al.*, Proximity Spin–Orbit Torque on a Two-Dimensional Magnet within van der Waals Heterostructure: Current-Driven Antiferromagnet-to-Ferromagnet Reversible Nonequilibrium Phase Transition in Bilayer CrI3, Nano Lett. **20**, 2288 (2020).
[20] J. Zhou *et al.*, Large tunneling magnetoresistance in VSe2/MoS2 magnetic tunnel junction, ACS Appl. Mater. Interfaces **11**, 17647 (2019).
[21] D. Apalkov *et al.*, Spin-transfer torque magnetic random access memory (STT-MRAM), ACM Journal on Emerging Technologies in Computing Systems (JETC) **9**, 1 (2013).
[22] E. Deng *et al.*, Low power magnetic full-adder based on spin transfer torque MRAM, IEEE Trans. Magn. **49**, 4982 (2013).
[23] W. Zhao *et al.*, Failure and reliability analysis of STT-MRAM, Microelectronics Reliability **52**, 1848 (2012).
[24] T. Seki *et al.*, High power radio frequency oscillation by spin transfer torque in a Co2MnSi layer: Experiment and macrospin simulation, J. Appl. Phys. **113**, 033907 (2013).
[25] W. Skowroński *et al.*, Spin-torque diode radio-frequency detector with voltage tuned resonance, Appl. Phys. Lett. **105**, 072409 (2014).
[26] A. Dussaux *et al.*, Large microwave generation from current-driven magnetic vortex oscillators in magnetic tunnel junctions, Nature communications **1**, 1 (2010).
[27] D. Fan *et al.*, STT-SNN: A spin-transfer-torque based soft-limiting non-linear neuron for low-power artificial neural networks, IEEE Transactions on Nanotechnology **14**, 1013 (2015).
[28] Z. Diao *et al.*, Spin-transfer torque switching in magnetic tunnel junctions and spin-transfer torque random access memory, J. Phys.: Condens. Matter **19**, 165209 (2007).
[29] J. Sun, and D. Ralph, Magnetoresistance and spin-transfer torque in magnetic tunnel junctions, J. Magn. Magn. Mater. **320**, 1227 (2008).
[30] B. K. Kaushik, and S. Verma, *Spin transfer torque based devices, circuits, and memory* (Artech House, 2016),
[31] L. Liu *et al.*, Spin-torque switching with the giant spin Hall effect of tantalum, Science **336**, 555 (2012).
[32] Y. Fan *et al.*, Magnetization switching through giant spin–orbit torque in a magnetically doped topological insulator heterostructure, Nature materials **13**, 699 (2014).
[33] J. R. Sánchez *et al.*, Spin-to-charge conversion using Rashba coupling at the interface between non-magnetic materials, Nature communications **4**, 1 (2013).
[34] J.-C. Rojas-Sánchez *et al.*, Spin to charge conversion at room temperature by spin pumping into a new type of topological insulator: α-Sn films, Phys. Rev. Lett. **116**, 096602 (2016).
[35] K. Dolui *et al.*, Proximity Spin-Orbit Torque on a Two-Dimensional Magnet within van der Waals Heterostructure: Current-Driven Antiferromagnet-to-Ferromagnet Reversible Nonequilibrium Phase Transition in Bilayer CrI3, Nano Lett **20**, 2288 (2020).
[36] K. Zollner *et al.*, Scattering-induced and highly tunable by gate damping-like spin-orbit torque in graphene doubly proximitized by two-dimensional magnet Cr2Ge2Te6 and monolayer WS2, Physical Review Research **2**, (2020).
[37] Y. Wang *et al.*, Topological surface states originated spin-orbit torques in Bi 2 Se 3, Phys. Rev. Lett. **114**, 257202 (2015).

[38] C. W. Smullen *et al.*, in *2011 IEEE 17th International Symposium on High Performance Computer Architecture* (IEEE, 2011), pp. 50.
[39] E. Chen *et al.*, Advances and future prospects of spin-transfer torque random access memory, IEEE Trans. Magn. **46**, 1873 (2010).
[40] S. Yuasa *et al.*, Giant room-temperature magnetoresistance in single-crystal Fe/MgO/Fe magnetic tunnel junctions, Nature materials **3**, 868 (2004).
[41] S. Ikeda *et al.*, A perpendicular-anisotropy CoFeB–MgO magnetic tunnel junction, Nature materials **9**, 721 (2010).
[42] M. Oogane *et al.*, Large tunnel magnetoresistance in magnetic tunnel junctions using Co2MnX (X= Al, Si) Heusler alloys, J. Phys. D: Appl. Phys. **39**, 834 (2006).
[43] J. Akerman *et al.*, Intrinsic reliability of AlOx-based magnetic tunnel junctions, IEEE Trans. Magn. **42**, 2661 (2006).
[44] V. Ilyasov *et al.*, Materials for spintronics: magnetic and transport properties of ultrathin (monolayer graphene)/MnO (001) and MnO (001) films, Journal of Modern Physics **2011**, (2011).
[45] P. Lv *et al.*, Half-metallicity in two-dimensional Co2Se3 monolayer with superior mechanical flexibility, 2D Materials **5**, 045026 (2018).
[46] J. Tuček *et al.*, Emerging chemical strategies for imprinting magnetism in graphene and related 2D materials for spintronic and biomedical applications, Chem. Soc. Rev. **47**, 3899 (2018).
[47] P. Hohenberg, and W. Kohn, Density functional theory (DFT), Phys. Rev **136**, B864 (1964).
[48] W. Kohn, and L. J. Sham, Self-consistent equations including exchange and correlation effects, Phys. Rev. **140**, A1133 (1965).
[49] W. Butler *et al.*, Spin-dependent tunneling conductance of Fe| MgO| Fe sandwiches, Physical Review B **63**, 054416 (2001).
[50] S. S. Parkin *et al.*, Giant tunnelling magnetoresistance at room temperature with MgO (100) tunnel barriers, Nature materials **3**, 862 (2004).
[51] A. Hirohata *et al.*, Review on Spintronics: Principles and Device Applications, Journal of Magnetism and Magnetic Materials, 166711 (2020).
[52] R. Fiederling *et al.*, Injection and detection of a spin-polarized current in a light-emitting diode, Nature **402**, 787 (1999).
[53] A. Joshua, and V. Venkataraman, Enhanced sensitivity in detection of Kerr rotation by double modulation and time averaging based on Allan variance, Review of Scientific Instruments **80**, 023908 (2009).
[54] Y. K. Kato *et al.*, Observation of the Spin Hall Effect in Semiconductors, Ence **306**, 1910 (2004).
[55] T. Jungwirth, J. Wunderlich, and K. Olejnik, Spin Hall effect devices, Nature Materials **11**, 382 (2012).
[56] Z. H. Xiong *et al.*, Giant magnetoresistance in organic spin-valves, Nature **427**, 821 (2004).
[57] LaBella, and P. V., Spatially Resolved Spin-Injection Probability for Gallium Arsenide, Science **292**, 1518 (2001).
[58] N. Tombros *et al.*, Electronic spin transport and spin precession in single graphene layers at room temperature, Nature, 571 (2007).
[59] B. Dlubak *et al.*, Highly efficient spin transport in epitaxial graphene on SiC, Nature Physics **8**, 557 (2012).
[60] E. I. Rashba, and A. L. Efros, Orbital Mechanisms of Electron-Spin Manipulation by an Electric Field, Physical Review Letters **91**, 126405 (2003).

[61] S. Prabhakar *et al.*, Manipulation of single electron spin in a GaAs quantum dot through the application of geometric phases: The Feynman disentangling technique, Physical Review B Condensed Matter **82**, 2460 (2010).
[62] J. A. H. Stotz *et al.*, Spin transport and manipulation by mobile potential dots in GaAs quantum wells, Physica E-low-dimensional Systems & Nanostructures **32**, 446 (2006).
[63] K. Y. Bliokh *et al.*, Spin–orbit interactions of light, Nature Photonics **9**, 796 (2015).
[64] Schewe, Phillip, and F., A single-spin transistor, Physics Today **55**, 9 (2002).
[65] M. Ciorga *et al.*, Collapse of the spin-singlett phase in quantum dots, Physical Review Letters **88**, 256804 (2002).
[66] S. Datta, and B. Das, Electronic analog of the electro-optic modulator, Applied Physics Letters **56**, 665 (1990).
[67] J. R. Petta *et al.*, Coherent manipulation of coupled electron spins in semiconductor quantum dots, Science **309**, 2180 (2005).
[68] R. Hanson, and D. D. Awschalom, Coherent manipulation of single spins in semiconductors, Nature **453**, 1043 (2008).
[69] P. Maurer *et al.*, Far-field optical imaging and manipulation of individual spins with nanoscale resolution, Nature Physics **6**, 912 (2010).
[70] W. Gao *et al.*, Coherent manipulation, measurement and entanglement of individual solid-state spins using optical fields, Nature Photonics **9**, 363 (2015).
[71] H. Mamin *et al.*, Magnetic resonance force microscopy of nuclear spins: Detection and manipulation of statistical polarization, Physical Review B **72**, 024413 (2005).
[72] C. Galland, and A. Imamoğlu, All-optical manipulation of electron spins in carbon-nanotube quantum dots, Phys. Rev. Lett. **101**, 157404 (2008).
[73] N. Linden *et al.*, Pulse sequences for NMR quantum computers: how to manipulate nuclear spins while freezing the motion of coupled neighbours, Chem. Phys. Lett. **305**, 28 (1999).
[74] Y. K. Kato, and D. D. Awschalom, Electrical manipulation of spins in nonmagnetic semiconductors, J. Phys. Soc. Jpn. **77**, 031006 (2008).
[75] M. Julliere, Tunneling between ferromagnetic films, Phys. Lett. A **54**, 225 (1975).
[76] J. M. De Teresa *et al.*, Role of Metal-Oxide Interface in Determining the Spin Polarization of Magnetic Tunnel Junctions, Science **286**, 507 (1999).
[77] T. Miyazaki, and N. Tezuka, Giant magnetic tunneling effect in Fe/Al2O3/Fe junction, J.magn.magn.mater **139**, 0 (1995).
[78] J. S. Moodera *et al.*, Large Magnetoresistance at Room Temperature in Ferromagnetic Thin Film Tunnel Junctions, Phys. Rev. Lett. **74**, 3273 (1995).
[79] H. X. Wei *et al.*, 80% tunneling magnetoresistance at room temperature for thin Al–O barrier magnetic tunnel junction with CoFeB as free and reference layers, J. Appl. Phys. **101**, 09B501 (2007).
[80] S. Ikeda *et al.*, Tunnel magnetoresistance of 604% at 300 K by suppression of Ta diffusion in CoFeB/MgO/CoFeB pseudo-spin-valves annealed at high temperature, Appl. Phys. Lett. **93**, 178 (2008).
[81] C. A. Ross, Patterned magnetic recording media, Annual Review of Materials Research **31**, 203 (2001).
[82] L. Pan, and D. B. Bogy, Heat-assisted magnetic recording, Nature Photonics **3**, 189 (2009).
[83] E. Hwang *et al.*, Interlaced magnetic recording, IEEE Trans. Magn. **53**, 1 (2016).
[84] C. Vogler *et al.*, Areal density optimizations for heat-assisted magnetic recording of high-density

media, J. Appl. Phys. **119**, 223903 (2016).

[85] S. Granz *et al.*, Perpendicular Interlaced Magnetic Recording, IEEE Trans. Magn. **55**, 1 (2019).

[86] A. Narahara, K. Ito, and T. Suemasu, Growth of ferromagnetic Fe4N epitaxial layers and a-axis-oriented Fe4N/MgO/Fe magnetic tunnel junction on MgO (0 0 1) substrates using molecular beam epitaxy, J. Cryst. Growth **311**, 1616 (2009).

[87] M. Guth *et al.*, Temperature dependence of transport properties in ZnS-based magnetic tunnel junctions, Journal of Magnetism & Magnetic Materials **240**, 152 (2002).

[88] T. Marukame *et al.*, High tunnel magnetoresistance in epitaxial Co2Cr0.6Fe0.4Al/MgO/CoFe tunnel junctions, IEEE Trans. Magn. **41**, 2603 (2005).

[89] S. Isogami, M. Tsunoda, and M. Takahashi, 30-nm scale fabrication of magnetic tunnel junctions using EB assisted CVD hard masks, IEEE Trans. Magn. **41**, 3607 (2005).

[90] M. Jílek Jr *et al.*, High-rate deposition of AlTiN and related coatings with dense morphology by central cylindrical direct current magnetron sputtering, Thin Solid Films **556**, 361 (2014).

[91] K. Tsuchiya, and S. Davies, Fabrication of TiNi shape memory alloy microactuators by ion beam sputter deposition, Nanotechnology **9**, 67 (1998).

[92] Y. You *et al.*, Electrical and optical study of ITO films on glass and polymer substrates prepared by DC magnetron sputtering type negative metal ion beam deposition, Mater. Chem. Phys. **107**, 444 (2008).

[93] S. Y. Lee *et al.*, Ion beam sputter deposited TiAlN films for metal-insulator-metal (Ba,Sr)TiO_3 capacitor application, Thin Solid Films **516**, p.7816 (2008).

[94] M. L. Piquemal-Banci *et al.*, Insulator-to-Metallic Spin-Filtering in 2D-Magnetic Tunnel Junctions Based on Hexagonal Boron Nitride, Acs Nano, acsnano.8b01354 (2018).

[95] G. Sasaki, S. Fujita, and A. Sasaki, Frequency dependence of flat‐band capacitance of metal/insulator/glow-discharge deposited hydrogenated amorphous silicon diodes, J. Appl. Phys. **55**, P.3183 (1984).

[96] C. Park *et al.*, Inverse magnetoresistance in magnetic tunnel junction with a plasma-oxidized Fe electrode and the effect of annealing on its transport properties, J. Appl. Phys. **97**, p.10C907.1 (2005).

[97] S. Yoshimura *et al.*, Oxidation process of Mg films by using high-concentration ozone for magnetic tunnel junctions, Journal of Magnetism & Magnetic Materials **312**, 176 (2007).

[98] H. Sun *et al.*, Deep-ultraviolet emitting AlGaN multiple quantum well graded-index separate-confinement heterostructures grown by MBE on SiC substrates, IEEE Photonics Journal **9**, 1 (2017).

[99] Z. Wang *et al.*, Tunneling spin valves based on Fe3GeTe2/hBN/Fe3GeTe2 van der Waals heterostructures, Nano letters **18**, 4303 (2018).

[100] A. Fert, and H. Jaffrès, Conditions for efficient spin injection from a ferromagnetic metal into a semiconductor, Physical Review B **64**, (2001).

[101] E. Y. Tsymbal *et al.*, Interface effects in spin-dependent tunneling, Progress in Materials Science **52**, 401 (2007).

[102] O. Wunnicke *et al.*, Effects of resonant interface states on tunneling magnetoresistance, Physical Review B **65**, 064425 (2002).

[103] B. Dlubak *et al.*, Are Al2O3 and MgO tunnel barriers suitable for spin injection in graphene?, Applied Physics Letters **97**, 092502 (2010).

[104] C. Józsa *et al.*, Linear scaling between momentum and spin scattering in graphene, Physical Review B **80**, 241403 (2009).

[105] B. Dlubak *et al.*, Homogeneous pinhole free 1 nm Al2O3 tunnel barriers on graphene, Applied Physics Letters **101**, 203104 (2012).
[106] T. Yamaguchi *et al.*, Electrical Spin Injection into Graphene through Monolayer Hexagonal Boron Nitride, Applied Physics Express **6**, 073001 (2013).
[107] K. D. Belashchenko, J. Velev, and E. Y. Tsymbal, Effect of interface states on spin-dependent tunneling in Fe / Mg O / Fe tunnel junctions, Physical Review B **72**, 140404 (2005).
[108] J. Zhang, X.-G. Zhang, and X. Han, Spinel oxides: Δ 1 spin-filter barrier for a class of magnetic tunnel junctions, Applied Physics Letters **100**, 222401 (2012).
[109] J. D. Burton *et al.*, Atomic and electronic structure of the CoFeB / MgO interface from first principles, Applied Physics Letters **89**, 142507 (2006).
[110] J. Koike *et al.*, Growth kinetics and thermal stability of a self-formed barrier layer at Cu-Mn / SiO2 interface, Journal of Applied Physics **102**, 043527 (2007).
[111] G.-X. Miao *et al.*, Inelastic tunneling spectroscopy of magnetic tunnel junctions based on CoFeB / MgO / CoFeB with Mg insertion layer, Journal of Applied Physics **99**, 08T305 (2006).
[112] D. D. Djayaprawira *et al.*, 230% room-temperature magnetoresistance in CoFeB / MgO / CoFeB magnetic tunnel junctions, Applied Physics Letters **86**, 092502 (2005).
[113] S. Ota *et al.*, CoFeB/MgO-based magnetic tunnel junction directly formed on a flexible substrate, Applied Physics Express **12**, 053001 (2019).
[114] W. Dexin *et al.*, 70% TMR at room temperature for SDT sandwich junctions with CoFeB as free and reference Layers, IEEE Transactions on Magnetics **40**, 2269 (2004).
[115] B. Dieny, and M. Chshiev, Perpendicular magnetic anisotropy at transition metal/oxide interfaces and applications, Reviews of Modern Physics **89**, 025008 (2017).
[116] R. Wood, Future hard disk drive systems, Journal of Magnetism and Magnetic Materials **321**, 555 (2009).
[117] A. Dankert *et al.*, Tunnel magnetoresistance with atomically thin two-dimensional hexagonal boron nitride barriers, Nano Research **8**, 1357 (2015).
[118] K. S. Novoselov *et al.*, Electric Field Effect in Atomically Thin Carbon Films, Science **306**, 666 (2004).
[119] J. Zhou *et al.*, 2DMatPedia, an open computational database of two-dimensional materials from top-down and bottom-up approaches, Scientific data **6**, 1 (2019).
[120] C. Gong, and X. Zhang, Two-dimensional magnetic crystals and emergent heterostructure devices, Science **363**, (2019).
[121] J. F. Dayen *et al.*, Two-dimensional van der Waals spinterfaces and magnetic-interfaces, Applied Physics Reviews **7**, 011303 (2020).
[122] V. Carteaux, F. Moussa, and M. Spiesser, 2D Ising-Like Ferromagnetic Behaviour for the Lamellar Cr2Si2Te6 Compound: A Neutron Scattering Investigation, Europhysics Letters **29**, 251 (1995).
[123] C. Gong *et al.*, Discovery of intrinsic ferromagnetism in two-dimensional van der Waals crystals, Nature **546**, 265 (2017).
[124] B. Huang *et al.*, Layer-dependent ferromagnetism in a van der Waals crystal down to the monolayer limit, Nature **546**, 270 (2017).
[125] V. P. Ningrum *et al.*, Recent Advances in Two-Dimensional Magnets: Physics and Devices towards Spintronic Applications, Research **2020**, 1 (2020).
[126] M. Wang *et al.*, Prospects and Opportunities of 2D van der Waals Magnetic Systems, Annalen der Physik **532**, 1900452 (2020).

[127] A. Banerjee *et al.*, Neutron scattering in the proximate quantum spin liquid α-RuCl3, Science **356**, 1055 (2017).
[128] W. Wang *et al.*, Spin-valve Effect in NiFe/MoS2/NiFe Junctions, Nano Letters **15**, 5261 (2015).
[129] E. Cobas *et al.*, Graphene As a Tunnel Barrier: Graphene-Based Magnetic Tunnel Junctions, Nano Letters **12**, 3000 (2012).
[130] E. Cobas *et al.*, Graphene-Based Magnetic Tunnel Junctions, IEEE Transactions on Magnetics **49**, 4343 (2013).
[131] M. Piquemal-Banci *et al.*, Magnetic tunnel junctions with monolayer hexagonal boron nitride tunnel barriers, Applied Physics Letters **108**, 102404 (2016).
[132] R. Urban, G. Woltersdorf, and B. Heinrich, Gilbert Damping in Single and Multilayer Ultrathin Films: Role of Interfaces in Nonlocal Spin Dynamics, Physical Review Letters **87**, 217204 (2001).
[133] T. Graf, C. Felser, and S. S. P. Parkin, Simple rules for the understanding of Heusler compounds, Progress in Solid State Chemistry **39**, 1 (2011).
[134] W. Rotjanapittayakul *et al.*, Spin injection and magnetoresistance in MoS 2 -based tunnel junctions using Fe 3 Si Heusler alloy electrodes, entific Reports **8**, 4779 (2018).
[135] A. Ionescu *et al.*, Structural, magnetic, electronic, and spin transport properties of epitaxial Fe3Si/GaAs(001), Phys.rev.b **71**, 094401(1 (2005).
[136] R. Meservey, and P. M. Tedrow, Spin-polarized electron tunneling, Physics Reports **238**, 173 (1994).
[137] H.-C. Wu *et al.*, Spin-dependent transport properties of Fe 3 O 4/MoS 2/Fe 3 O 4 junctions, Scientific reports **5**, 15984 (2015).
[138] H. Zhang *et al.*, Magnetoresistance in Co/2D MoS 2/Co and Ni/2D MoS 2/Ni junctions, Physical Chemistry Chemical Physics **18**, 16367 (2016).
[139] M. Galbiati *et al.*, Path to Overcome Material and Fundamental Obstacles in Spin Valves Based on Mo S 2 and Other Transition-Metal Dichalcogenides, Physical Review Applied **12**, 044022 (2019).
[140] J.-J. Jin *et al.*, Magnetotransport in a zigzag monolayer MoS2 nanoribbon with ferromagnetic electrodes, Phys. Lett. A **383**, 125852 (2019).
[141] J. Liu *et al.*, Exfoliating biocompatible ferromagnetic Cr-trihalide monolayers, PCCP **18**, 8777 (2016).
[142] Z. Yan *et al.*, Significant tunneling magnetoresistance and excellent spin filtering effect in CrI3-based van der Waals magnetic tunnel junctions, PCCP **22**, (2020).
[143] T. Song *et al.*, Giant tunneling magnetoresistance in spin-filter van der Waals heterostructures, Science **360**, 1214 (2018).
[144] Z. Z. Lin, and X. Chen, Ultrathin Scattering Spin Filter and Magnetic Tunnel Junction Implemented by Ferromagnetic 2D van der Waals Material, Advanced Electronic Materials **6**, 1900968 (2020).
[145] Z. Wang *et al.*, Very large tunneling magnetoresistance in layered magnetic semiconductor CrI 3, Nature communications **9**, 1 (2018).
[146] D. R. Klein *et al.*, Probing magnetism in 2D van der Waals crystalline insulators via electron tunneling, Science **360**, 1218 (2018).
[147] Y. Deng *et al.*, Gate-tunable room-temperature ferromagnetism in two-dimensional Fe 3 GeTe 2, Nature **563**, 94 (2018).
[148] X. Li *et al.*, Spin-dependent transport in van der Waals magnetic tunnel junctions with Fe3GeTe2 electrodes, Nano Lett. **19**, 5133 (2019).

[149] L. Zhang *et al.*, Perfect Spin Filtering Effect on Fe3GeTe2-Based Van der Waals Magnetic Tunnel Junctions, The Journal of Physical Chemistry C **124**, 27429 (2020).
[150] L. Pan *et al.*, Two-dimensional XSe2 (X= Mn, V) based magnetic tunneling junctions with high Curie temperature, Chinese Physics B **28**, 107504 (2019).
[151] W. Rotjanapittayakul *et al.*, Spin injection and magnetoresistance in MoS 2-based tunnel junctions using Fe 3 Si Heusler alloy electrodes, Scientific reports **8**, 1 (2018).
[152] L. Song *et al.*, Realizing robust half-metallic transport with chemically modified graphene nanoribbons, Carbon **141**, 676 (2019).
[153] L. B. Frechette, C. Dellago, and P. L. Geissler, Consequences of lattice mismatch for phase equilibrium in heterostructured solids, Physical review letters **123**, 135701 (2019).
[154] Z. Yang *et al.*, A Fermi-Level-Pinning-Free 1D Electrical Contact at the Intrinsic 2D MoS2–Metal Junction, Advanced Materials **31**, 1808231 (2019).
[155] W. Hou *et al.*, Strain-based room-temperature non-volatile MoTe2 ferroelectric phase change transistor, Nature Nanotechnology **14**, 1 (2019).
[156] M. Zhou, H. Jin, and Y. Xing, In-Plane Dual-Gated Spin-Valve Device Based on the Zigzag Graphene Nanoribbon, Physical Review Applied **13**, 044006 (2020).
[157] Y. Zhang *et al.*, Tuning the electrical conductivity of Ti2CO2 MXene by varying the layer thickness and applying strains, The Journal of Physical Chemistry C **123**, 6802 (2019).
[158] X. Zhang *et al.*, High and anisotropic carrier mobility in experimentally possible Ti 2 CO 2 (MXene) monolayers and nanoribbons, Nanoscale **7**, 16020 (2015).
[159] Y. Zhou *et al.*, Electronic and transport properties of Ti2CO2 MXene nanoribbons, The Journal of Physical Chemistry C **120**, 17143 (2016).
[160] E. Balcı, U. n. O. z. Akkuş, and S. Berber, High TMR in MXene-Based Mn2CF2/Ti2CO2/Mn2CF2 Magnetic Tunneling Junction, ACS Appl. Mater. Interfaces **11**, 3609 (2018).
[161] Ü. Ö. Akkuş, E. Balcı, and S. Berber, Device characteristics of Ti2CT2 MXene-based field-effect transistor, Superlattices Microstruct. **140**, 106433 (2020).
[162] W. Qiu *et al.*, Spin-dependent resonant tunneling and magnetoresistance in Ni/graphene/h-BN/graphene/Ni van der Waals heterostructures, J. Magn. Magn. Mater. **476**, 622 (2019).
[163] M. R. Sahoo *et al.*, First-principles study of a vertical spin switch in atomic scale two-dimensional platform, J. Magn. Magn. Mater. **484**, 462 (2019).
[164] Q. Wu *et al.*, Efficient spin injection into graphene through a tunnel barrier: overcoming the spin-conductance mismatch, Physical Review Applied **2**, 044008 (2014).
[165] D. J. O'Hara *et al.*, Room temperature intrinsic ferromagnetism in epitaxial manganese selenide films in the monolayer limit, Nano Lett. **18**, 3125 (2018).
[166] A. Dankert *et al.*, Tunnel magnetoresistance with atomically thin two-dimensional hexagonal boron nitride barriers, Nano Research **8**, 1357 (2015).
[167] P. Asshoff *et al.*, Magnetoresistance of vertical Co-graphene-NiFe junctions controlled by charge transfer and proximity-induced spin splitting in graphene, 2D Materials **4**, 031004 (2017).
[168] M. Piquemal-Banci *et al.*, Magnetic tunnel junctions with monolayer hexagonal boron nitride tunnel barriers, Appl. Phys. Lett. **108**, 102404 (2016).
[169] M. I. Piquemal-Banci *et al.*, Insulator-to-metallic spin-filtering in 2D-magnetic tunnel junctions based on hexagonal boron nitride, ACS nano **12**, 4712 (2018).
[170] L. Pan *et al.*, Large tunneling magnetoresistance in magnetic tunneling junctions based on two-dimensional CrX 3 (X= Br, I) monolayers, Nanoscale **10**, 22196 (2018).

[171] L. V. Begunovich *et al.*, Triple VTe2/graphene/VTe2 heterostructures as perspective magnetic tunnel junctions, Appl. Surf. Sci. **510**, 145315 (2020).
[172] J. J. Heath *et al.*, Role of quantum confinement and interlayer coupling in CrI 3-graphene magnetic tunnel junctions, Physical Review B **101**, 195439 (2020).
[173] M. Chen *et al.*, Nonequilibrium spin injection in monolayer black phosphorus, PCCP **18**, 1601 (2016).
[174] T. Song *et al.*, Switching 2D magnetic states via pressure tuning of layer stacking, Nature materials, 1 (2019).
[175] A. Avsar *et al.*, van der Waals Bonded Co/h-BN Contacts to Ultrathin Black Phosphorus Devices, Nano Letters **17**, 5361 (2017).
[176] M. Bonilla *et al.*, Strong room-temperature ferromagnetism in VSe 2 monolayers on van der Waals substrates, Nature nanotechnology **13**, 289 (2018).
[177] J. Yang *et al.*, Rationally designed high-performance spin filter based on two-dimensional half-metal Cr2NO2, Matter **1**, 1304 (2019).
[178] N. Nishimura *et al.*, Magnetic tunnel junction device with perpendicular magnetization films for high-density magnetic random access memory, J. Appl. Phys. **91**, 5246 (2002).
[179] Y. Ke, K. Xia, and H. Guo, Disorder scattering in magnetic tunnel junctions: Theory of nonequilibrium vertex correction, Phys. Rev. Lett. **100**, 166805 (2008).
[180] A. A. Starikov *et al.*, Calculating the transport properties of magnetic materials from first principles including thermal and alloy disorder, noncollinearity, and spin-orbit coupling, Physical Review B **97**, 214415 (2018).
[181] Z. Jiang *et al.*, Screening and design of novel 2D ferromagnetic materials with high curie temperature above room temperature, ACS Appl. Mater. Interfaces **10**, 39032 (2018).